\documentclass[submitting,onecolumn]{nst}

\usepackage[utf8]{inputenc}
\usepackage[T2A,T1]{fontenc}
\usepackage[russian,english]{babel}
\usepackage{graphicx}
\usepackage[singlelinecheck=false]{caption}
\usepackage{subcaption}
\usepackage{tikz}
\usepackage{amsmath}
\usepackage{dcolumn}
\usepackage{bm}
\usepackage{etoolbox}
\usepackage{listings}
\usepackage{upgreek}
\usepackage{csquotes}
\usepackage{epstopdf}
\usepackage{mhchem}
\usepackage{siunitx}
\usepackage{hyperref}
\usepackage{cleveref}
\usepackage{xcolor}
\usepackage{textcomp}
\usepackage{ulem,fancyvrb}
\usepackage{booktabs}
\begin{document}

\title{
A REST-based Monte Carlo Simulation and Data Analysis Framework for the N$\nu$DEx Experiment
}

\author{Siyuan Huang}
\email[Corresponding author, ]{huangsiyuan22@mails.ucas.ac.cn}
\affiliation{
State Key Laboratory of Heavy Ion Science and Technology, Institute of Modern Physics, Chinese Academy of Sciences, Lanzhou 730000, China
}

\author{Hao Qiu}
\affiliation{
State Key Laboratory of Heavy Ion Science and Technology, Institute of Modern Physics, Chinese Academy of Sciences, Lanzhou 730000, China
}

\author{Shaobo Wang}
\affiliation{SPEIT (SJTU-Paris Elite Institute of Technology), Shanghai Jiao Tong University, Shanghai 200240, China}
\affiliation{Shanghai Jiao Tong University Sichuan Research Institute, Chengdu 610213, China}
\affiliation{INPAC and School of Physics and Astronomy, Shanghai Jiao Tong University, MOE Key Lab for Particle Physics, Astrophysics and Cosmology, Shanghai Key Laboratory for Particle Physics and Cosmology, Shanghai 200240, China}
\affiliation{Jinping Deep Underground Frontier Science and Dark Matter Key Laboratory of Sichuan Province, Xichang 615000, China}

\author{Tao Li}
\email[Corresponding author, ]{taoli@sjtu.edu.cn}
\affiliation{SPEIT (SJTU-Paris Elite Institute of Technology), Shanghai Jiao Tong University, Shanghai 200240, China}
\affiliation{Shanghai Jiao Tong University Sichuan Research Institute, Chengdu 610213, China}

\begin{abstract}
The N$\nu$DEx experiment is designed to search for neutrinoless double-beta decay ($0\nu\beta\beta$) of $\mathrm{^{82}Se}$ and is planned to be deployed at the China Jinping Underground Laboratory (CJPL). It employs a high-pressure time projection chamber (TPC) filled with $\mathrm{^{82}SeF_6}$ gas and a Topmetal pixelated charge readout system. The high $Q_{\beta\beta}$ value of $\mathrm{^{82}Se}$, together with the track reconstruction capability of a gaseous TPC, provides strong potential for background suppression.
In this work, we develop a full simulation and reconstruction framework for the N$\nu$DEx experiment based on the REST (Rare Event Searches Toolkit) software framework. To meet the detector-specific requirements of N$\nu$DEx, customized simulation processes are implemented to describe the detector geometry, electron drift and diffusion, the Topmetal readout plane, signal formation, and electronics response. These developments enable a complete software description of the detector response from energy deposition to digitized waveforms.
A waveform-fitting-based reconstruction method is further applied to the simulated signals to extract the deposited charge and reconstruct the event energy. For simulated signal events, the reconstructed energy resolution reaches a full width at half maximum (FWHM) of less than 1\%, satisfying the design requirement of the experiment. The resulting simulation and reconstruction chain provides a basis for future studies of detector response, background rejection, event classification, and optimization of the N$\nu$DEx detector design.
\end{abstract}


\maketitle
\nolinenumbers

\section{Introduction}
\label{sec:introduction}

Neutrino oscillation and the implied nonzero neutrino mass constitute one of the most compelling pieces of evidence for physics beyond the Standard Model (SM). This discovery has been confirmed by several experiments, including Super-Kamiokande~\cite{PhysRevLett.81.1562} and SNO~\cite{PhysRevLett.89.011301}. Oscillation experiments can measure mass-squared differences but cannot determine whether the neutrino is a Dirac or a Majorana particle. As early as 1937, Ettore Majorana proposed that electrically neutral fermions could be their own antiparticles~\cite{Majorana1937}. If neutrinos are Majorana fermions, a process known as neutrinoless double-beta decay ($0\nu\beta\beta$) becomes possible, providing a direct experimental test of this fundamental question~\cite{Dolinski2019}.

In the Standard Model-allowed two-neutrino double-beta decay ($2\nu\beta\beta$), two neutrons in a candidate nucleus simultaneously convert into two protons while emitting two electrons and two electron antineutrinos. This process has been observed in several isotopes, including $^{82}\mathrm{Se}$, $^{76}\mathrm{Ge}$, $^{130}\mathrm{Te}$, and $^{136}\mathrm{Xe}$. If neutrinos are Majorana fermions, $0\nu\beta\beta$ can occur without the emission of electron antineutrinos. According to the ``black-box theorem,'' the observation of $0\nu\beta\beta$ would directly establish the Majorana nature of neutrinos~\cite{PhysRevD.25.2951}. Moreover, such a discovery would provide support for the seesaw mechanism, which naturally explains why neutrino masses are far smaller than those of the charged leptons~\cite{PhysRevLett.44.912}. Observation of $0\nu\beta\beta$ would also signify lepton-number violation, with far-reaching implications for neutrino physics and related fields. Despite nearly 80 years of experimental effort, however, no definitive evidence for $0\nu\beta\beta$ has been obtained. Several major experiments are currently collecting data or under construction: GERDA~\cite{PhysRevLett.125.252502}, MAJORANA~\cite{PhysRevLett.120.132502}, LEGEND and CUPID~\cite{CUPID2025}, KamLAND-Zen~\cite{PhysRevLett.130.051801}, PandaX~\cite{Ni_2019}, and JUNO~\cite{Zhao_2017} are all pursuing this discovery.

\begin{figure}[!htb]
    \centering
    \includegraphics[width=0.45\textwidth]{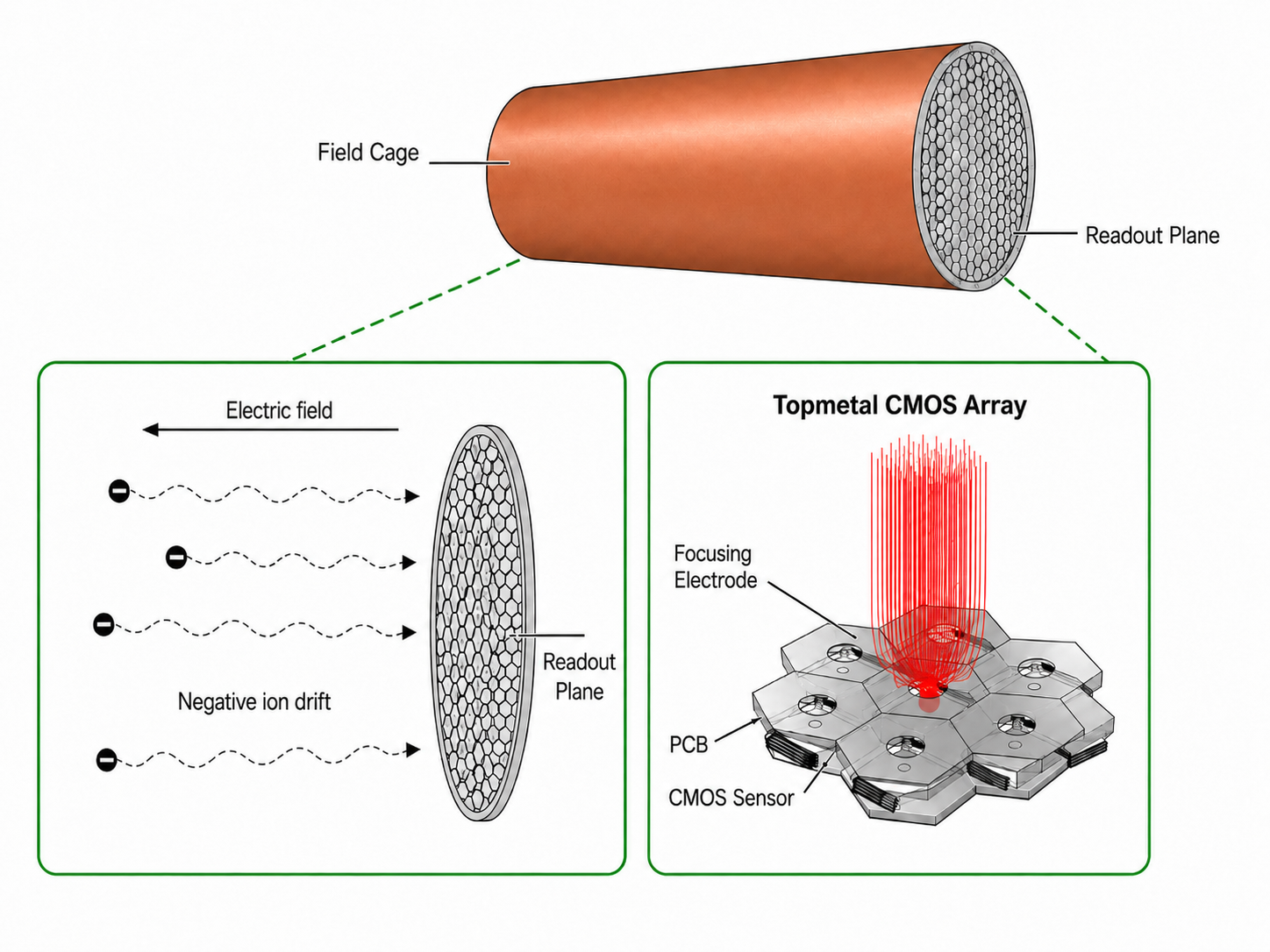}
    \caption{ Schematic view of the N$\nu$DEx-100 detector concept. The detector uses high-pressure $\mathrm{^{82}SeF_6}$ gas as both the source and the detection medium, with Topmetal silicon sensors for charge readout.}
    \label{fig:NvDEx100}
\end{figure}

The N$\nu$DEx experiment, which aims to search for $0\nu\beta\beta$ of $^{82}\mathrm{Se}$, was first proposed in 2018~\cite{Nygren_2018}. It employs a high-pressure gaseous time projection chamber (TPC) filled with $\mathrm{^{82}SeF_6}$ as the working medium and uses Topmetal silicon sensors for charge readout. As the first phase of the N$\nu$DEx program, N$\nu$DEx-100 has been proposed as a 100-kg-scale detector to be deployed at the China Jinping Underground Laboratory (CJPL)~\cite{Cao_2024_NvDEx100}. Its conceptual design report~\cite{Cao_2024_NvDEx100} presents the overall detector scheme, major subsystems, background estimation, and projected sensitivity, providing the technical basis for the subsequent development of the N$\nu$DEx experiment. Figure~\ref{fig:NvDEx100} shows a schematic view of the N$\nu$DEx-100 detector concept.

N$\nu$DEx takes advantage of the high $Q_{\beta\beta}$ value of $^{82}\mathrm{Se}$ and the particle-track measurement capability of the gaseous TPC to achieve high sensitivity to $0\nu\beta\beta$. The $Q_{\beta\beta}$ value of $^{82}\mathrm{Se}$, 2.996~MeV, lies above most natural radioactive backgrounds and is higher than the decay energies of many isotopes used in leading experiments. For example, the intensity of natural radioactive $\gamma$ backgrounds near the $Q_{\beta\beta}$ value of $^{82}\mathrm{Se}$ is more than two orders of magnitude lower than that near the $Q_{\beta\beta}$ value of $^{136}\mathrm{Xe}$ (2.458~MeV). In a gaseous TPC, moreover, the ionization tracks of the two electrons from double-beta decay can be precisely reconstructed. Each track typically exhibits a Bragg peak near its end point, providing a powerful means of distinguishing signal from background and thus enhancing the experimental sensitivity.

In this work, we develop a REST-based full simulation and reconstruction chain for the N$\nu$DEx experiment. The simulation implements the high-pressure $\mathrm{^{82}SeF_6}$ gaseous TPC geometry, relevant physical processes, the Topmetal pixelated charge readout, and the electronics response. A waveform-fitting-based reconstruction method is then applied to the simulated signals to extract the deposited charge and reconstruct the event energy. Using this framework, we evaluate the performance of the full simulation and establish a foundation for detector-response studies and reconstruction optimization in N$\nu$DEx.

This paper is organized as follows. Section~\ref{sec:rest_framework} introduces the REST software framework. Section~\ref{sec:simulation} describes the implementation of the full simulation workflow. Section~\ref{sec:reconstruction} discusses the event reconstruction method and results. Section~\ref{sec:conclusion} summarizes the work and provides an outlook.

\section{The REST Framework}
\label{sec:rest_framework}

\begin{figure*}[!htb]
    \centering
    \includegraphics[width=0.75\textwidth]{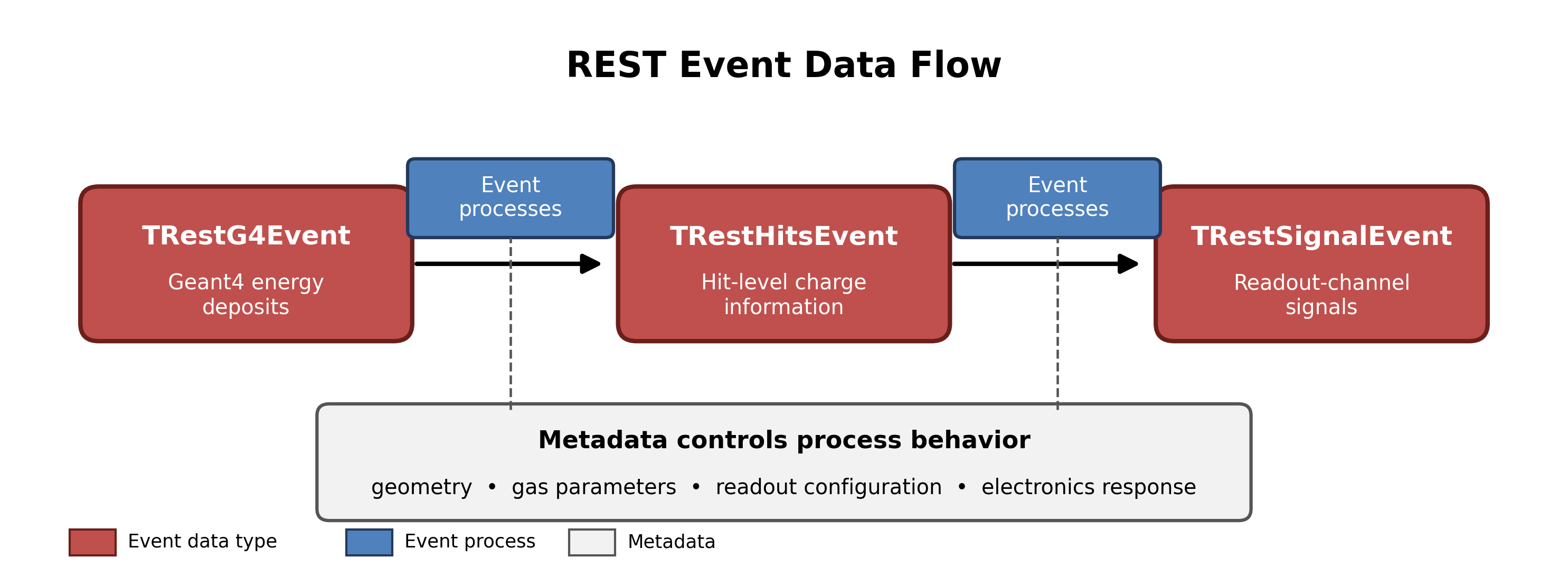}
    \caption{ Simplified data-flow structure in the REST framework. Detector information is represented by different event data types: \textit{TRestG4Event}, \textit{TRestHitsEvent}, and \textit{TRestSignalEvent}. Transformations between these event types are performed by event processes controlled by the corresponding metadata.}
    \label{fig:REST_structure}
\end{figure*}

REST (Rare Event Searches Toolkit) is a ROOT-based software framework designed to provide a general-purpose toolkit for experimental particle physics, supporting data acquisition, simulation, data analysis, and detector response studies~\cite{ALTENMULLER2022108281}. The framework handles both Monte Carlo data generation and experimental data reconstruction and emphasizes code traceability and result validation. REST enables the same event processing pipeline to be used for Monte Carlo and experimental data, facilitating direct comparisons between simulation and measurement.

Figure~\ref{fig:REST_structure} illustrates the basic data processing workflow for Monte Carlo events in REST. Detector information at different simulation stages is stored in distinct event data types. The \textit{TRestG4Event} type stores energy deposits and particle information produced by Geant4-based simulations. The \textit{TRestHitsEvent} type represents hit-level information, such as the spatial distribution of charge after detector-response processes. The \textit{TRestSignalEvent} type stores the signals on readout channels generated from the hit information.

The transformation between these event types is performed by REST event processes. Each process applies a specific detector or electronics effect to the input event and produces a new event type or an updated event content. Their behavior is controlled by metadata objects that define the relevant detector properties, such as gas parameters, readout geometry, and electronics response. This modular structure allows the simulation and reconstruction chain to be constructed by connecting event processes in sequence. The implementation for N$\nu$DEx is described in the following sections.

\section{Simulation Workflow for N$\nu$DEx}
\label{sec:simulation}

This section describes the simulation workflow developed for the N$\nu$DEx experiment within the REST framework. The simulation models the detector response from the initial particle interaction in the high-pressure $\mathrm{SeF_6}$ gas to the formation of readout signals on the Topmetal pixelated plane. Three main components are implemented: the detector geometry, the Topmetal readout structure, and the physical processes that govern particle transport, negative-ion drift and diffusion, signal formation, and electronics response.

We first construct the detector geometry according to the conceptual design of the N$\nu$DEx-100 detector, including the shielding structures, high-pressure gas chamber, inner copper shield, insulation layer, and active gas volume. We then model the Topmetal pixelated readout plane to describe the segmentation of the charge collection plane and the mapping between drifting ions and readout pixels. Finally, we implement the physical processes to generate primary events, convert energy deposits into ionization signals, simulate negative-ion drift and diffusion in $\mathrm{SeF_6}$ gas, and produce digitized waveforms after the electronics response. The following subsections provide the details.

\subsection{Detector Geometry}

\begin{figure*}[htbp]
    \centering

    \begin{subfigure}[b]{0.45\textwidth}
        \centering
        \includegraphics[width=\textwidth]{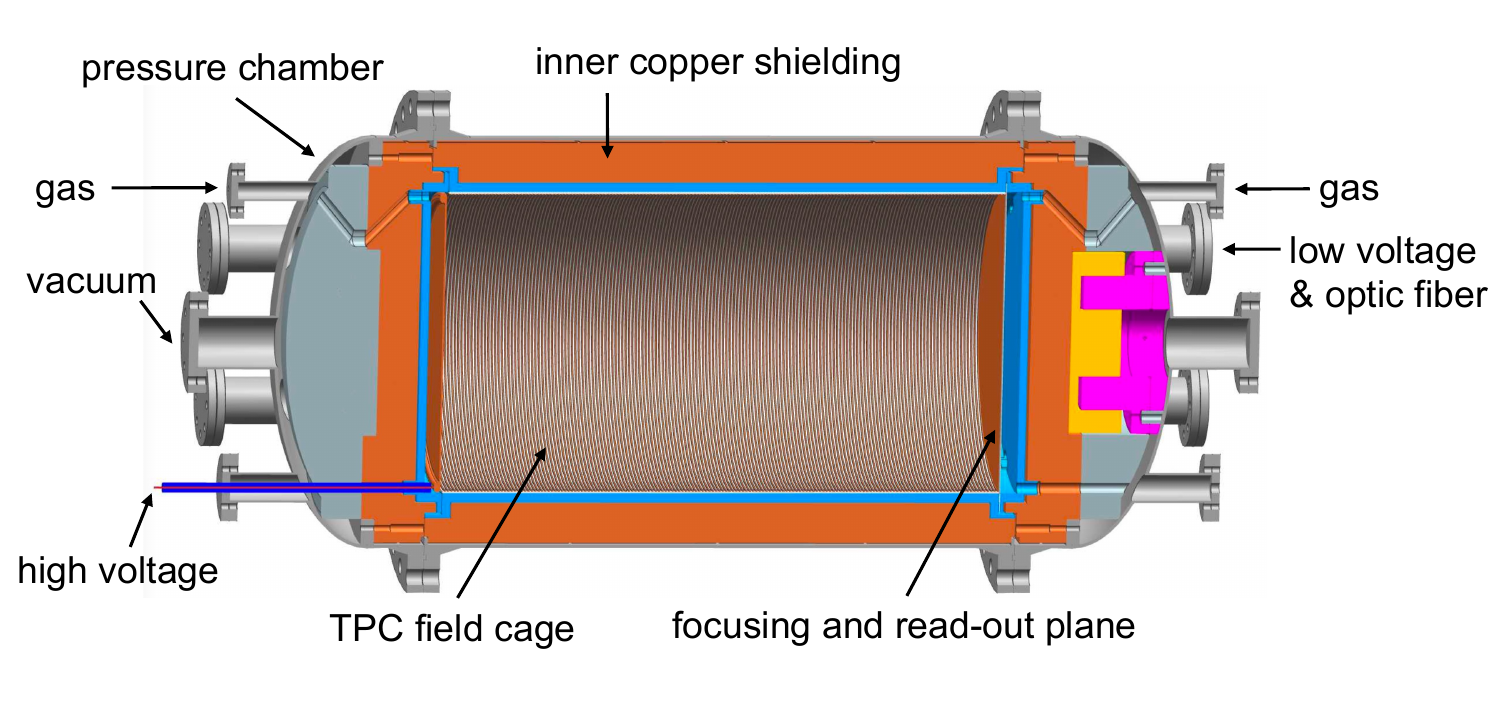}
        \caption{}
        \label{fig:NvDEx100_design}
    \end{subfigure}
    \hfill
    \begin{subfigure}[b]{0.45\textwidth}
        \centering
        \includegraphics[width=\textwidth]{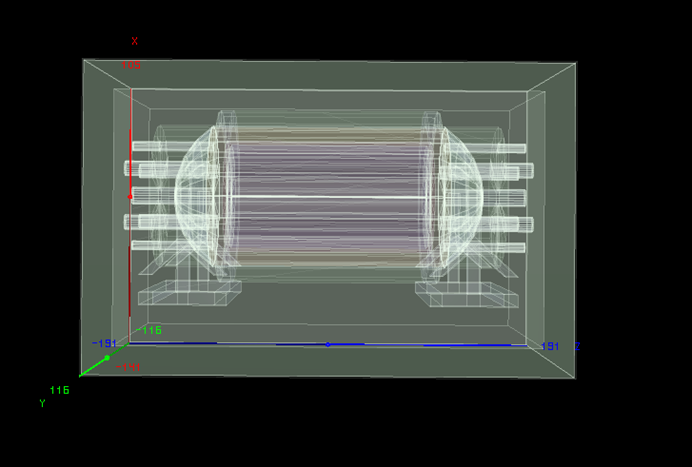}
        \caption{}
        \label{fig:Monolithic_structure}
    \end{subfigure}

    \caption{Detector design of the N$\nu$DEx-100 experiment and its geometry implementation in the REST framework.
    (a) Conceptual design of the N$\nu$DEx-100 detector, reproduced from Ref.~\cite{Cao_2024_NvDEx100}.
    (b) Corresponding geometry model constructed in the REST framework for detector-response simulation.}
    \label{fig:detector_design}
\end{figure*}

The detector geometry follows the conceptual design of the N$\nu$DEx-100 detector~\cite{Cao_2024_NvDEx100}, as illustrated in Fig.~\ref{fig:detector_design}. The overall setup consists of an external shielding system and a high-pressure gaseous TPC located inside the shielding volume. From outside to inside, the main structures include a high-density polyethylene (HDPE) shield, a lead shield, the high-pressure gas chamber, an inner low-background oxygen-free copper shield, an electrical insulation layer, and the active $\mathrm{SeF_6}$ gas volume.

The external shielding system is designed to suppress environmental radiation backgrounds. It consists of a 20-cm-thick HDPE shield as the outermost layer and a 20-cm-thick lead shield inside it for the attenuation of external $\gamma$ rays. The remaining space between the lead shield and the high-pressure gas chamber is filled with HDPE blocks, providing additional shielding.

The high-pressure stainless-steel vessel consisting of a cylindrical barrel, two  endcaps, flanges, and service pipes. The barrel has an inner diameter of 120~cm, a length of 160~cm, and a thickness of 1~cm. The  endcaps are placed at both ends of the barrel, while the flanges and service pipes represent the main structural components of the chamber. These elements are included in the geometry model to provide a realistic description of material distribution around the active volume.

Inside the pressure vessel, a 12-cm-thick low-background oxygen-free copper shield reduces radioactive backgrounds from the pressure vessel and the surrounding environment. The copper shield has an inner diameter of 96~cm and a length of 176~cm. A 20-mm-thick polyoxymethylene (POM) cylindrical layer serves as an electrical insulation layer, with an inner diameter of 89~cm and a length of 169~cm. The active gas region is a cylindrical volume filled with 1.0~MPa $\mathrm{SeF_6}$ gas, which serves as both the source material and the detection medium.

All geometry components are described using a GDML file and imported into the REST framework. The corresponding materials --- HDPE, lead, stainless steel, oxygen-free copper, POM, and $\mathrm{SeF_6}$ gas --- are assigned to the relevant volumes. This geometry model provides the basis for the subsequent simulation of particle interactions, charge transport, and detector response.

\subsection{Topmetal Pixelated Readout Plane}

The N$\nu$DEx detector employs a segmented charge-readout plane based on Topmetal-S sensors to measure the ionization charges produced in high-pressure $\mathrm{SeF_6}$ gas. Topmetal-S is a single-pixel CMOS direct-charge sensor specifically developed for the N$\nu$DEx experiment. Fabricated in a 130-nm CMOS process, the sensor integrates an exposed hexagonal charge-collection electrode (CCE), a low-noise charge-sensitive amplifier (CSA), and peripheral readout circuitry. The CCE is approximately 1~mm in diameter and is surrounded by a guard ring to assist charge collection~\cite{Yang_2024_TopmetalS,Liang_2024_TopmetalS}. Owing to its low-noise front-end electronics, Topmetal-S is capable of directly detecting the small charge signals produced by drifting electrons and ions without relying on gas-avalanche amplification. Its capability to detect electrons as well as both negative and positive ions in gas has been experimentally demonstrated~\cite{Liang_2024_TopmetalS}. This direct and highly segmented charge-readout scheme is particularly suitable for N$\nu$DEx, where detailed event-topology reconstruction is required for background discrimination in the $0\nu\beta\beta$ search.

The readout plane is located at one end of the TPC sensitive volume, and the maximum drift distance in the detector model used in this work is 169~cm. In the actual N$\nu$DEx design, an array of Topmetal-S sensors forms the charge-readout plane, together with a focusing structure that guides the drifting ions toward the charge-collection electrodes~\cite{Cao_2024_NvDEx100}. For the detector simulation presented in this work, the readout plane is represented by 32 modules, each divided into $16\times16=256$ effective hexagonal readout cells, resulting in a total of 8192 simulated readout channels.

\begin{figure*}[htbp]
    \centering

    \begin{subfigure}[t]{0.45\textwidth}
        \centering
        \includegraphics[height=4cm,keepaspectratio]{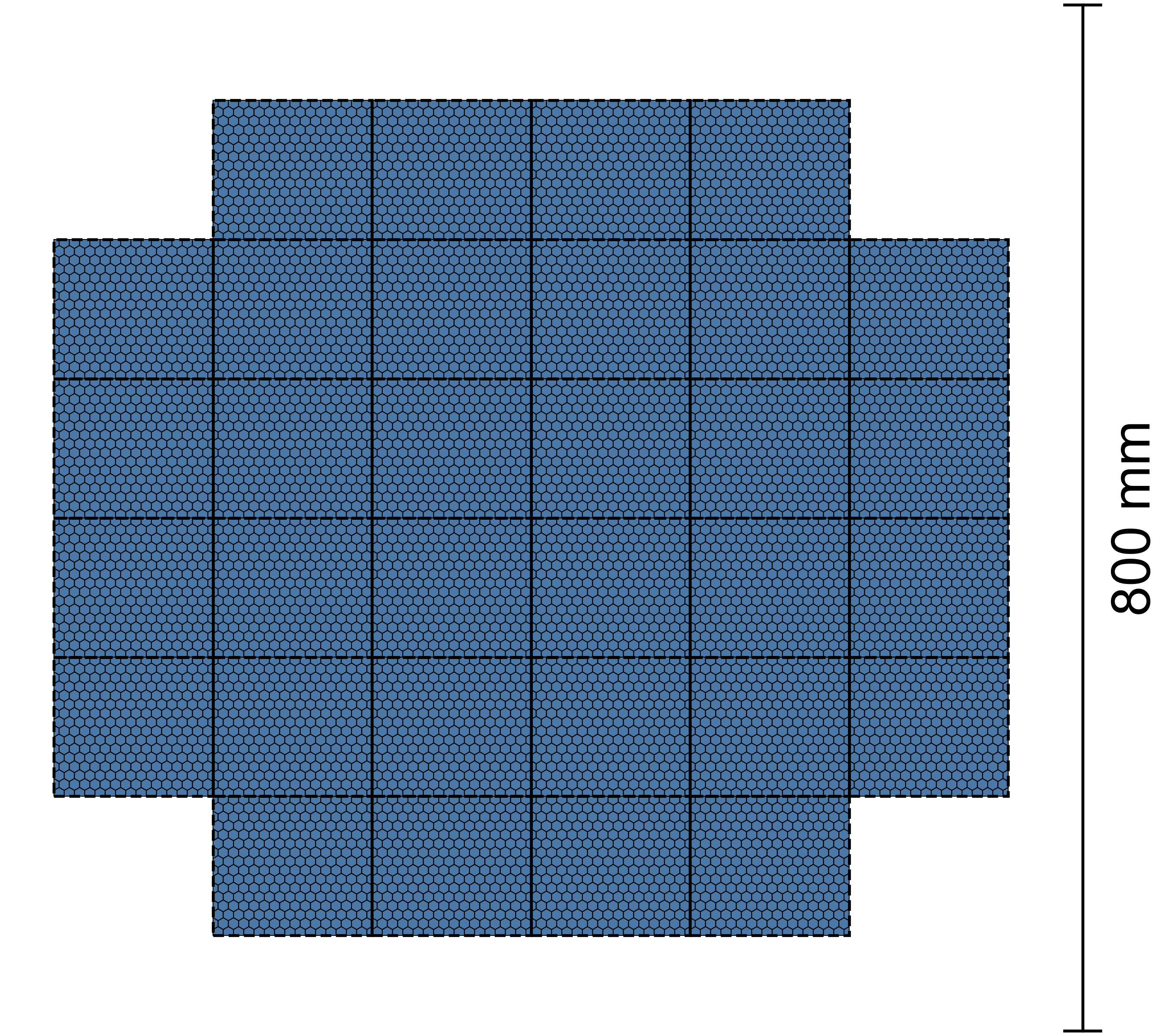}
        \caption{}
        \label{fig:readout_plane_simulated}
    \end{subfigure}
    \hfill
    \begin{subfigure}[t]{0.45\textwidth}
        \centering
        \includegraphics[height=4cm,keepaspectratio]{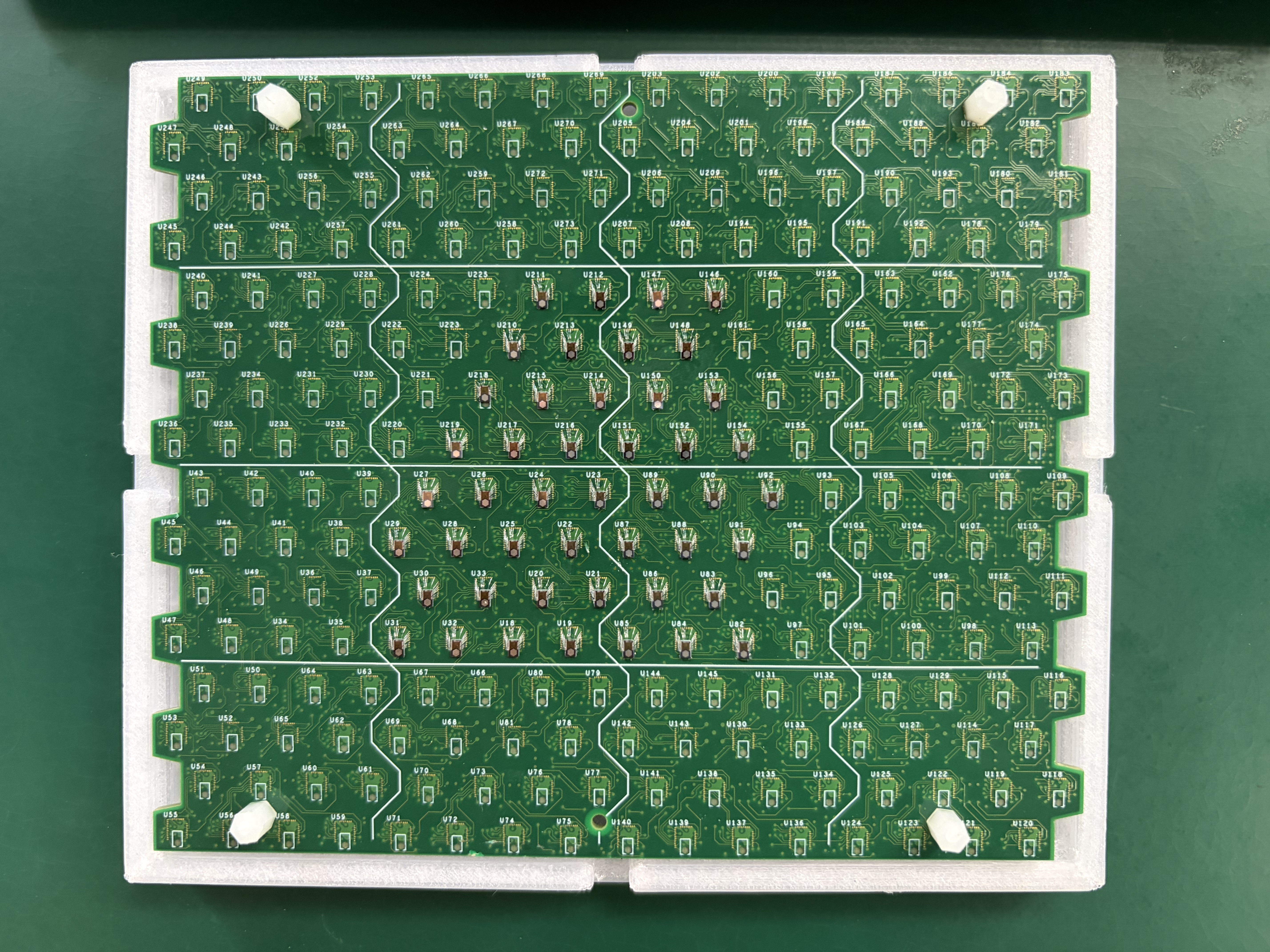}
        \caption{}
        \label{fig:readout_module_real}
    \end{subfigure}

    \caption{Readout-module design and simulation setup for the N$\nu$DEx detector.
    (a) Full readout plane implemented in the simulation, composed of multiple readout modules.
    (b) Construction scheme of a single readout module.}
    \label{fig:readout_plane_setup}
\end{figure*}

The effective readout cells used in the simulation are arranged on a regular hexagonal grid with a center-to-center spacing of 8~mm between adjacent cells. This corresponds to a hexagon side length of $8/\sqrt{3}\approx4.62$~mm. Each hexagonal cell represents one effective readout channel in the detector-response simulation and should not be interpreted as the physical charge-collection electrode of an individual Topmetal-S sensor. This simplified segmentation is used to map the transported ionization charges onto discrete readout channels for subsequent signal formation and event reconstruction. The simulated readout plane and the corresponding readout-module design are shown in Fig.~\ref{fig:readout_plane_setup}.

\subsection{Physical Processes for Detector Response}

\begin{figure*}[!htb]
    \centering
    \includegraphics[width=0.85\textwidth]{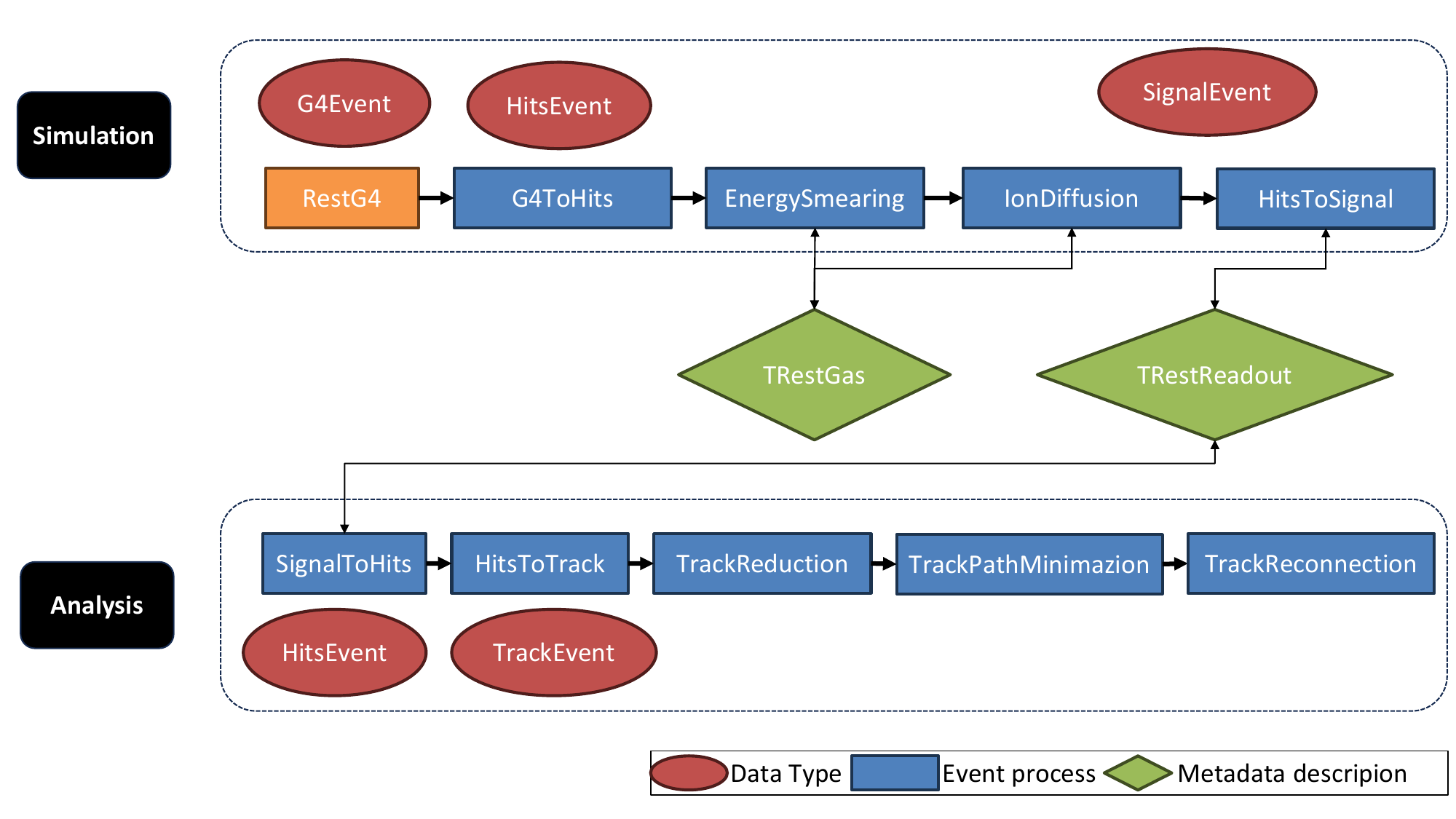}
    \caption{Full simulation and reconstruction workflow implemented in the REST framework for the N$\nu$DEx detector. The upper part shows the simulation chain from Geant4 energy deposits to detector signals, including hit conversion, ion drift and diffusion, and readout response. The lower part shows the reconstruction chain from detector signals to reconstructed hits and tracks. The corresponding event types, event processes, and metadata objects are shown in the diagram.}
    \label{fig:simulation_workflow}
\end{figure*}

The complete simulation and reconstruction workflow implemented for the N$\nu$DEx experiment is shown in Fig.~\ref{fig:simulation_workflow}. The workflow starts from the Geant4-based event generation and proceeds through the conversion of energy deposits into hit information, ion drift and diffusion, readout response, signal formation, and subsequent reconstruction processes.

Primary events are generated using the \textsc{Decay0} event generator, which is integrated into the REST framework. The $0\nu\beta\beta$ decay of $^{82}\mathrm{Se}$ is simulated with the $Q_{\beta\beta}$ value set to 2.996~MeV. Decay0 provides the initial kinematics of the two emitted electrons, which are then passed to Geant4 for particle transport within the detector geometry. The physics list is configured to include the standard electromagnetic processes for the energy range relevant to $^{82}\mathrm{Se}$ $\beta\beta$ decay. Geant4 records the position, time, and amount of energy deposited at each step in the active gas volume, constituting the primary ionization signal. The deposited energy is converted into a number of ionization electrons using the average ionization energy $W = 32$~eV for $\mathrm{SeF_6}$ gas, sampled from a Gaussian distribution whose width is determined by the Fano factor $F = 0.19$. Approximately $9.35\times10^{4}$ primary electrons are produced per $0\nu\beta\beta$ decay at the $Q_{\beta\beta}$ energy.

Since $\mathrm{SeF_6}$ is a strongly electronegative gas, free electrons produced by ionization are rapidly captured by nearby molecules, so charge transport is dominated by negative ions rather than free electrons. The main drifting species are $\mathrm{SeF_5^-}$ and $\mathrm{SeF_6^-}$, which have different masses and thus different drift velocities under the same electric field. This difference produces a distinct double-peak structure in the readout waveforms, providing timing information for reconstructing the drift distance. The REST drift process models the full charge transport: for each ionization electron, it computes the drift time of each ion species using their respective drift velocities, applies longitudinal and transverse diffusion by smearing the arrival position and time, and assigns the charge to the readout pixel whose hexagonal cell contains the arrival position.

The drift velocity and diffusion coefficients are determined by the reduced mobility $\mu_0$, the electric field strength $E$, the gas pressure $P$, and the temperature $T$. For $\mathrm{SeF_6^-}$, the measured reduced mobility of $\mathrm{SF_6^-}$ in $\mathrm{SF_6}$ gas is used as a reference~\cite{Phan_2017_SF6}. For other $\mathrm{SeF}_n^{\pm}$ ions, the reduced mobility is estimated using an approximate mass-scaling relation based on a simplified ion--molecule collision model (see Appendix~\ref{app:mobility_scaling}). The drift parameters used in the simulation are summarized in Table~\ref{tab:drift_parameters}.

\begin{table*}[!htb]
\caption{Drift parameters for different ions.}
\label{tab:drift_parameters}
\centering
\begin{tabular*}{\textwidth}{@{\extracolsep{\fill}} lcccccccc}
\toprule
Ion & $\mathrm{SF}_6^{-}$ & $\mathrm{SF}_5^{-}$ & $\mathrm{SeF}_6^{-}$ & $\mathrm{SeF}_5^{-}$ & $\mathrm{SeF}_4^{-}$ & $\mathrm{SeF}_3^{-}$ & $\mathrm{SeF}_2^{-}$ & $\mathrm{SeF}^{-}$ \\
\midrule
$A$ & 146 & 127 & 196 & 177 & 158 & 139 & 120 & 101 \\
$\mu_0~(\mathrm{cm^2\,V^{-1}\,s^{-1}})$ & 0.54 & 0.59 & 0.46 & 0.50 & 0.54 & 0.59 & 0.66 & 0.74 \\
\bottomrule
\end{tabular*}
\end{table*}

The drift velocity $v_d$ of the ions in the TPC is given by
\begin{equation}
v_d = \frac{1~\mathrm{atm}}{P} \, \frac{T}{273~\mathrm{K}} \, \mu_0 E,
\label{eq:vd}
\end{equation}
where $P$ is the gas pressure, $T$ is the temperature, $\mu_0$ is the reduced mobility, and $E$ is the electric field strength. In the TPC, $P = 10$~atm, $T = 293.15$~K, and $E = 437$~V/m. Substituting the values from Table~\ref{tab:drift_parameters} into Eq.~\eqref{eq:vd} gives
giving
\begin{equation}
\begin{cases}
v_d(\mathrm{SeF_5^{-}}) = 23.63~\mathrm{cm/s},\\
v_d(\mathrm{SeF_6^{-}}) = 21.89~\mathrm{cm/s}.
\end{cases}
\label{eq:vd_both}
\end{equation}

The longitudinal diffusion coefficient $D_L$ is given by the Einstein relation:
\begin{equation}
D_L = \frac{\mu_0 k T}{e (P / 1~\mathrm{atm})},
\label{eq:Dz}
\end{equation}
where $k$ is the Boltzmann constant and $e$ is the elementary charge. The positional uncertainty due to diffusion follows a Gaussian distribution whose standard deviation depends on the drift distance $L$:
\begin{equation}
\sigma_L = \sqrt{\frac{2 D_L L}{v_d}}.
\label{eq:sigma_z}
\end{equation}

For an ionization point with energy loss $E_d$, the total number of ions $n$ follows a Gaussian distribution with mean $E_d / W$ and variance $F \cdot E_d / W$:
\begin{equation}
n \sim N\!\left(\frac{E_d}{W},\; \frac{F \cdot E_d}{W}\right),
\label{eq:n}
\end{equation}
where the number ratio of $\mathrm{SeF_5^-}$ ions to $\mathrm{SeF_6^-}$ ions is set to 1:2 based on the dissociation branching ratio of $\mathrm{SeF_6^-}$. After the ion number and species are determined, a positional uncertainty sampled from $N(0, \sigma_z^2)$ is added to the original coordinates of each ion, completing the simulation of ion drift and diffusion.

After reaching the readout plane, the diffused ions are mapped onto the hexagonal readout grid according to their transverse positions. Each ion is assigned to the corresponding readout cell, and the charges collected within the same cell are combined to form the response of that channel. In this way, the continuous spatial distribution of the drifting charges is converted into a discrete distribution of activated readout channels for subsequent signal formation and event reconstruction.

\begin{figure}[!htb]
    \centering

    \begin{subfigure}[t]{0.45\textwidth}
        \centering
        \includegraphics[width=\textwidth]{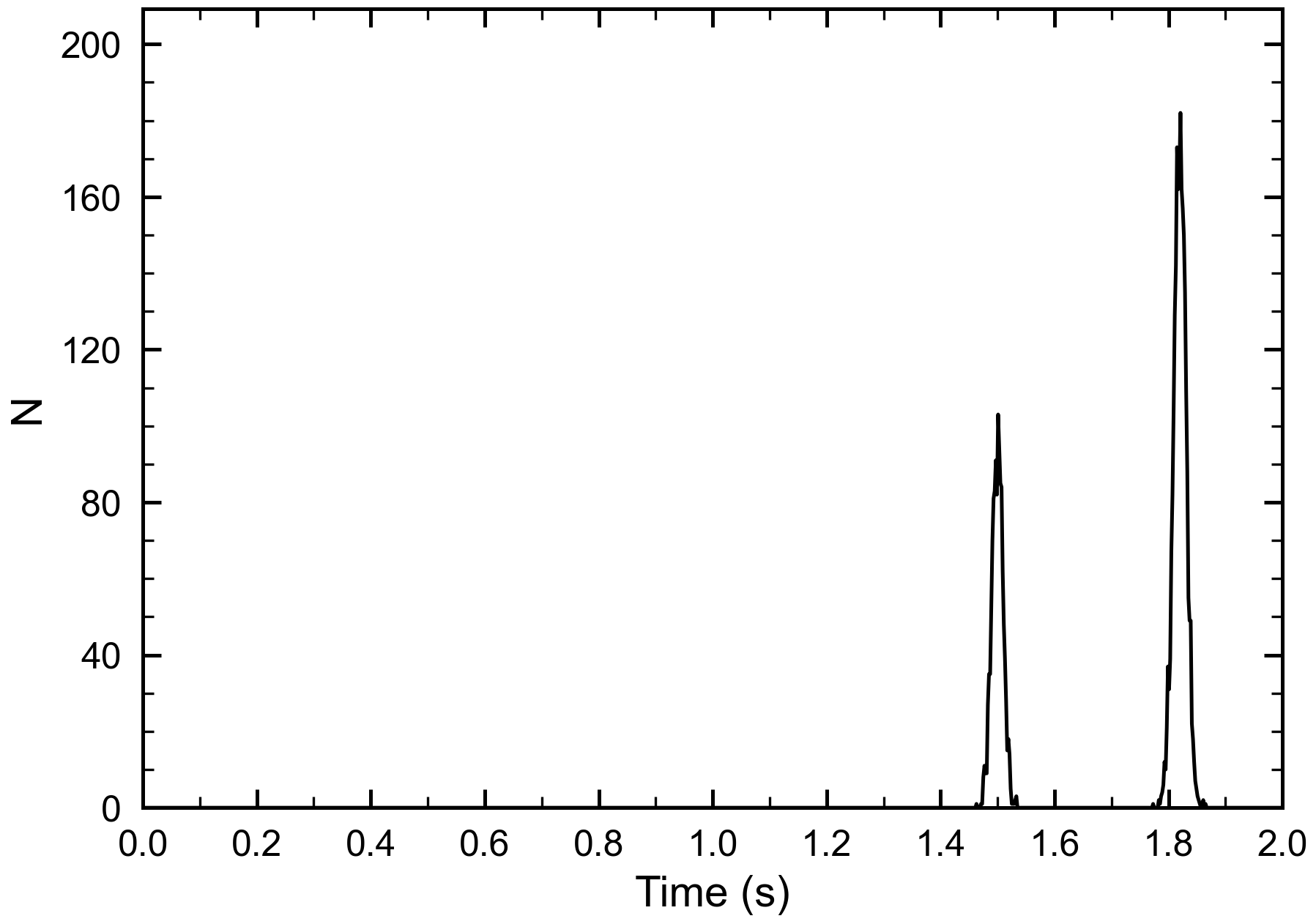}
        \caption{}
        \label{fig:charge_arrival}
    \end{subfigure}
    \hfill
    \begin{subfigure}[t]{0.45\textwidth}
        \centering
        \includegraphics[width=\textwidth]{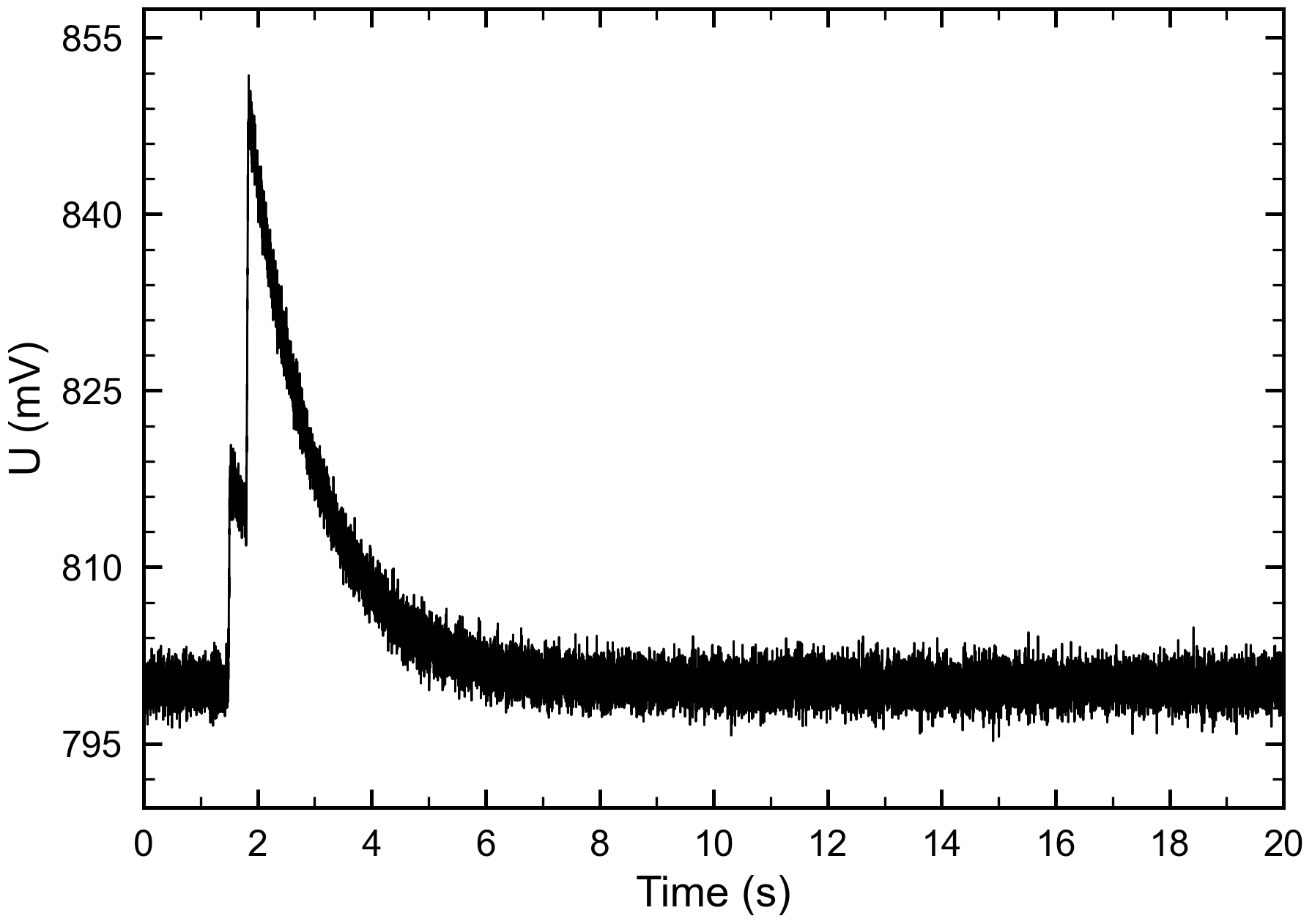}
        \caption{}
        \label{fig:electronics_response}
    \end{subfigure}

    \caption{Simulation of the readout-electronics response.
    (a) Charge-arrival-time distribution of the two ion species at the readout plane.
    (b) Corresponding output waveform obtained after convolution with the electronics response function and the addition of electronic noise.}
    \label{fig:signal_response}
\end{figure}

The voltage amplitude $U$ of a readout channel is related to the collected charge $Q$ through $U = Q/C_{\mathrm{fb}}$, where $C_{\mathrm{fb}}$ is the feedback capacitance of the charge-sensitive amplifier. The charge-arrival-time distribution obtained from the ion transport simulation is converted into an output waveform by convolving it with the response function of the readout electronics. The characteristic rise time of the response is approximately 50~ms. To provide sufficient sampling of the rising edge, a sampling interval of 4~ms (250~Hz) is used. For a drift distance of approximately 160~cm and an ion drift velocity of about 20~cm/s, the arrival-time difference can extend up to approximately 8~s. Considering the characteristic decay time of the electronics response, approximately 8~s, a total acquisition window of 20~s is adopted, corresponding to 5000 sampling points. Finally, Gaussian electronic noise with an RMS level equivalent to 45~$e^{-}$ is added to the simulated waveform. Figure~\ref{fig:signal_response} illustrates the charge-arrival-time distribution and the corresponding waveform after applying the electronics response and noise.

\begin{figure*}[!htb]
    \centering

    \begin{subfigure}[t]{0.31\textwidth}
        \centering
        \includegraphics[width=\textwidth]{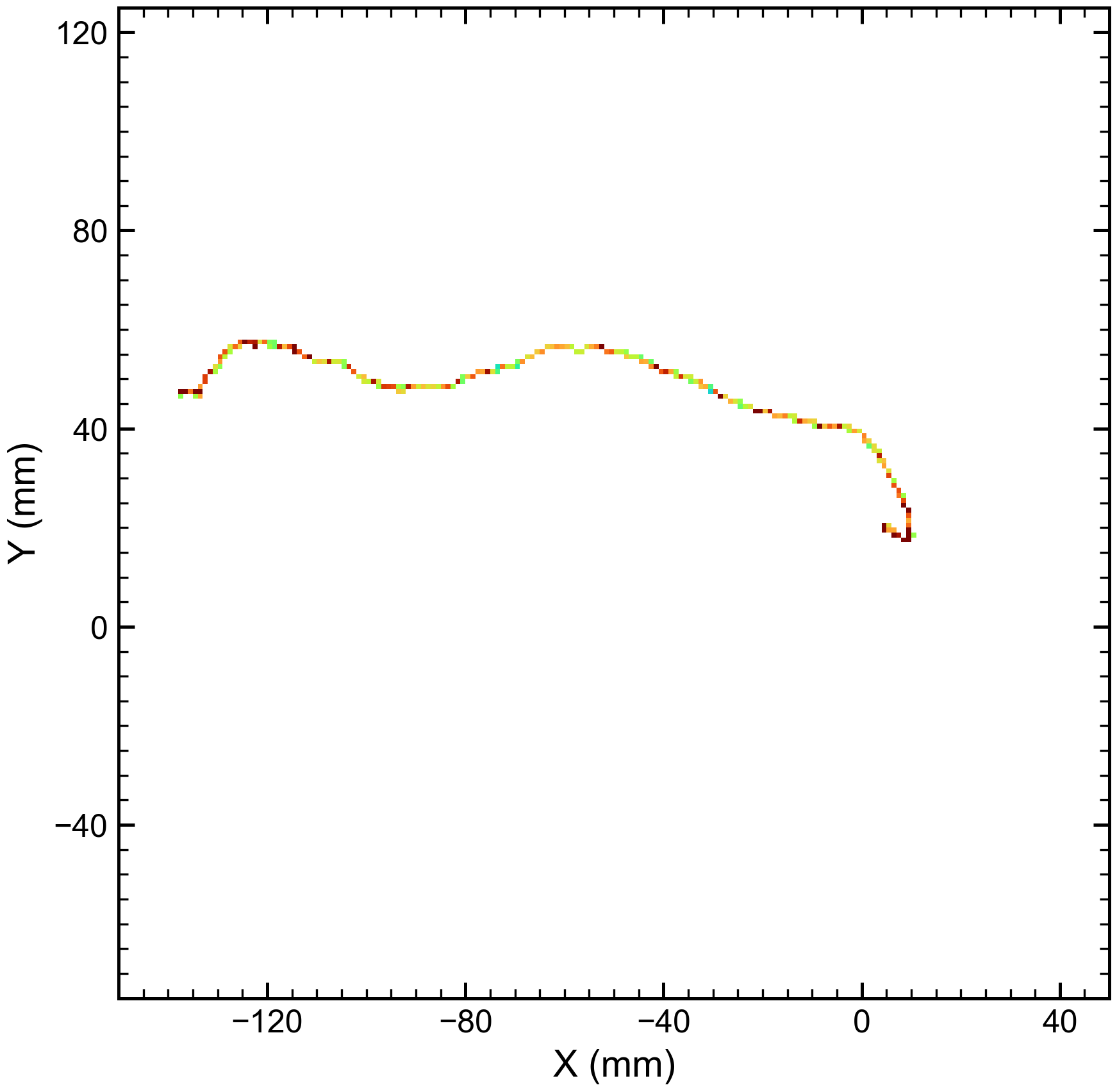}
        \caption{}
        \label{fig:original_xy}
    \end{subfigure}
    \hfill
    \begin{subfigure}[t]{0.31\textwidth}
        \centering
        \includegraphics[width=\textwidth]{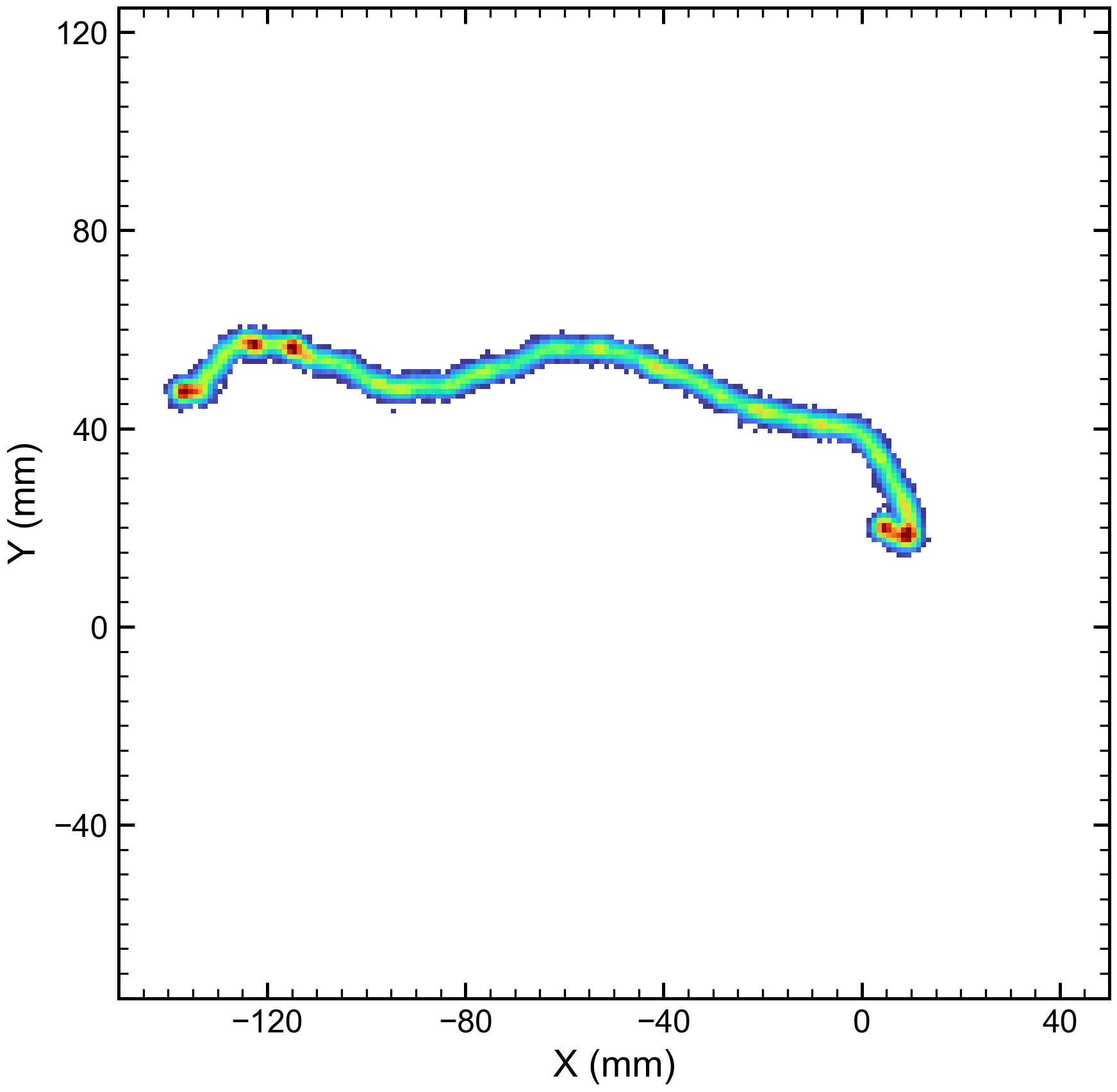}
        \caption{}
        \label{fig:diffusion_xy}
    \end{subfigure}
    \hfill
    \begin{subfigure}[t]{0.31\textwidth}
        \centering
        \includegraphics[width=\textwidth]{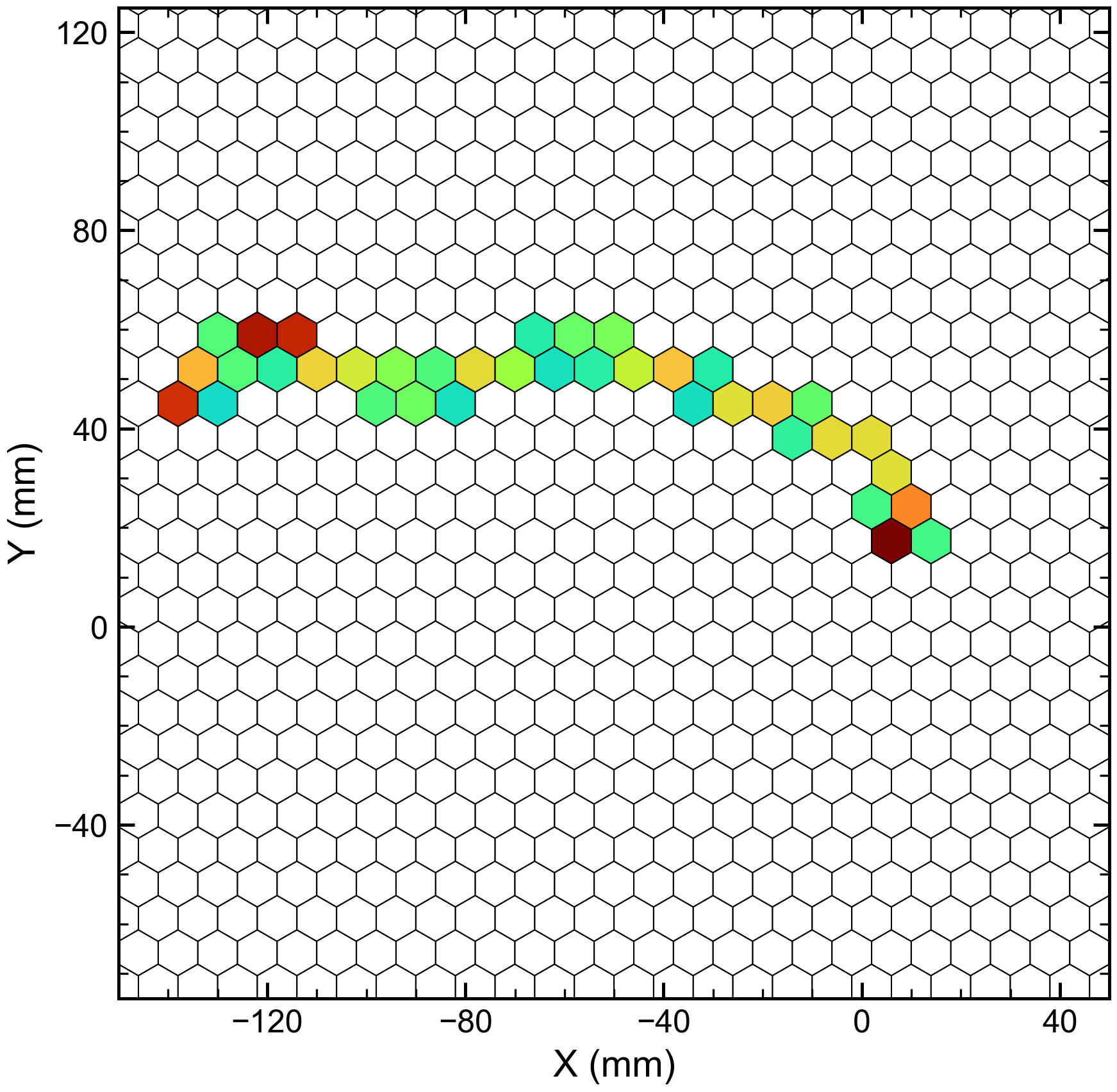}
        \caption{}
        \label{fig:rebuild_xy_hexagon}
    \end{subfigure}

    \caption{Evolution of a simulated track in the XY projection.
    (a) Original track generated by Geant4.
    (b) Track after ion diffusion.
    (c) Corresponding distribution of responding readout pixels on the hexagonal Topmetal readout plane.}
    \label{fig:track_xy_response}
\end{figure*}

To illustrate the detector spatial response, Fig.~\ref{fig:track_xy_response} shows the evolution of a simulated track in the XY projection, including the original Geant4 track, the charge distribution after ion diffusion, and the corresponding response on the hexagonal readout plane. As expected, ion diffusion broadens the spatial distribution of the charges, while the finite segmentation of the hexagonal readout plane further discretizes the track pattern. These effects are consistent with the designed spatial response of the detector.

\section{Event Reconstruction}
\label{sec:reconstruction}

This section describes the reconstruction of event properties from the simulated chip signals. We apply a waveform-fitting method to each readout channel, then combine the results to produce the reconstructed event track and energy.

The relationship between the voltage amplitude $U$ on a readout pixel and the collected charge $Q$ is $U = Q/C_{\mathrm{fb}}$, where $C_{\mathrm{fb}}$ is the feedback capacitance of the charge-sensitive amplifier in the Topmetal pixel. The charge $Q$ is related to the number of collected electrons $N$ by $N = Q/e$, where $e$ is the elementary charge. The deposited energy is then obtained by multiplying the total number of collected electrons by the average ionization energy $W = 32$~eV.

\subsection{Waveform Fitting Method}

To reduce the impact of waveform fluctuations and make full use of timing and amplitude information, we apply a waveform-fitting method rather than directly extracting peak positions and heights.

The fitting function is constructed by convolving the ion-arrival distribution with the electronics response of the chip. The ion-arrival distribution is modeled by a double-Gaussian function, corresponding to the arrivals of $\mathrm{SeF_5^-}$ and $\mathrm{SeF_6^-}$ ions. The electronics response function describes the shaping effect of the readout circuit. The complete fitting function is
\begin{equation}
f(t)=g(t)\ast h(t),
\label{eq:fit_function}
\end{equation}
where $g(t)$ is the double-Gaussian ion-arrival distribution:
\begin{equation}
g(t)=\frac{A_1}{\sqrt{2\pi}\sigma_1}e^{-0.5\left(\frac{t-t_1}{\sigma_1}\right)^2}
     +\frac{A_2}{\sqrt{2\pi}\sigma_2}e^{-0.5\left(\frac{t-t_2}{\sigma_2}\right)^2}.
\label{eq:double_gaussian_function}
\end{equation}
Here, $t_1$ and $t_2$ are the arrival times of $\mathrm{SeF_5^-}$ and $\mathrm{SeF_6^-}$, respectively, and $A_1$ and $A_2$ represent the corresponding signal amplitudes. The widths $\sigma_1$ and $\sigma_2$ describe the temporal spread of the ion clouds caused by longitudinal diffusion during drift.

The chip response function is
\begin{equation}
h(t)=\left(1-e^{-t/\tau_1}\right)e^{-t/\tau_2},
\label{eq:response_function}
\end{equation}
where $\tau_1$ and $\tau_2$ are the characteristic time constants of the chip response. The parameter $\tau_1$ describes the rise-time behavior of the shaped signal, and $\tau_2$ describes the exponential decay after charge collection. Convolving $g(t)$ with $h(t)$ yields a fitting function that captures both the double-arrival structure and the broadening introduced by the electronics.

The diffusion-induced temporal widths are determined by the longitudinal diffusion coefficient and the drift velocity of each ion species:
\begin{align}
\sigma_1 &= \frac{\sqrt{2D_L(\mathrm{SeF_5^-})L/v_d(\mathrm{SeF_5^-})}}{v_d(\mathrm{SeF_5^-})},
\label{eq:sigma1} \\
\sigma_2 &= \frac{\sqrt{2D_L(\mathrm{SeF_6^-})L/v_d(\mathrm{SeF_6^-})}}{v_d(\mathrm{SeF_6^-})},
\label{eq:sigma2}
\end{align}
where $D_L$ is the longitudinal diffusion coefficient, $v_d$ is the drift velocity, and $L$ is the drift distance. These equations reflect the fact that a longer drift distance results in a broader arrival-time distribution due to diffusion.

The drift distance is obtained from the arrival-time difference between the two ion species. Because both originate from the same ionization point but drift at different velocities, the time separation between the two peaks is related to the common drift distance:
\begin{equation}
L = \frac{(t_2 - t_1)\,v_d(\mathrm{SeF_5^-})\,v_d(\mathrm{SeF_6^-})}
          {v_d(\mathrm{SeF_5^-}) - v_d(\mathrm{SeF_6^-})}.
\label{eq:L}
\end{equation}

\begin{figure}[!htb]
    \centering
    \includegraphics[width=0.55\textwidth]{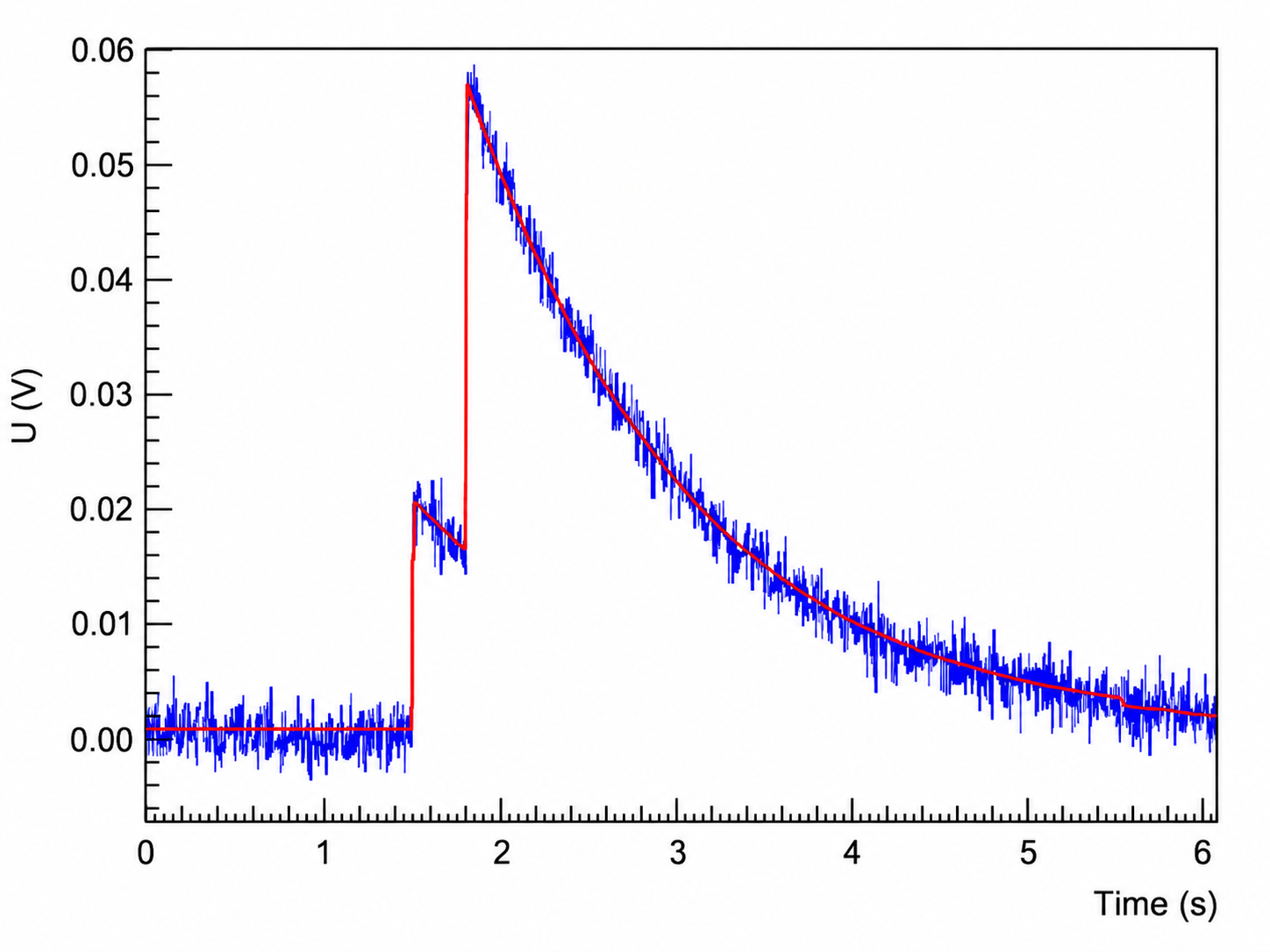}
    \caption{Example of waveform fitting for a double-peak signal induced by the arrivals of $\mathrm{SeF_5^-}$ and $\mathrm{SeF_6^-}$ ions.}
    \label{fig:fitting_signal}
\end{figure}

In the fitting procedure, $t_1$, $t_2$, $A_1$, and $A_2$ are treated as free parameters, while the diffusion widths are determined by the drift distance and the ion transport parameters. The waveform fitting is performed using ROOT. Figure~\ref{fig:fitting_signal} shows an example fitting result.

\subsection{Fitting Optimization}

At a fixed channel noise level of 45~$e^{-}$, signals with small amplitudes are submerged in noise, causing the fit to fail. We therefore require a threshold $S_t$: signals with amplitudes above $S_t$ are retained, while those below are discarded.

\begin{figure}[!htb]
    \centering
    \includegraphics[width=0.55\textwidth]{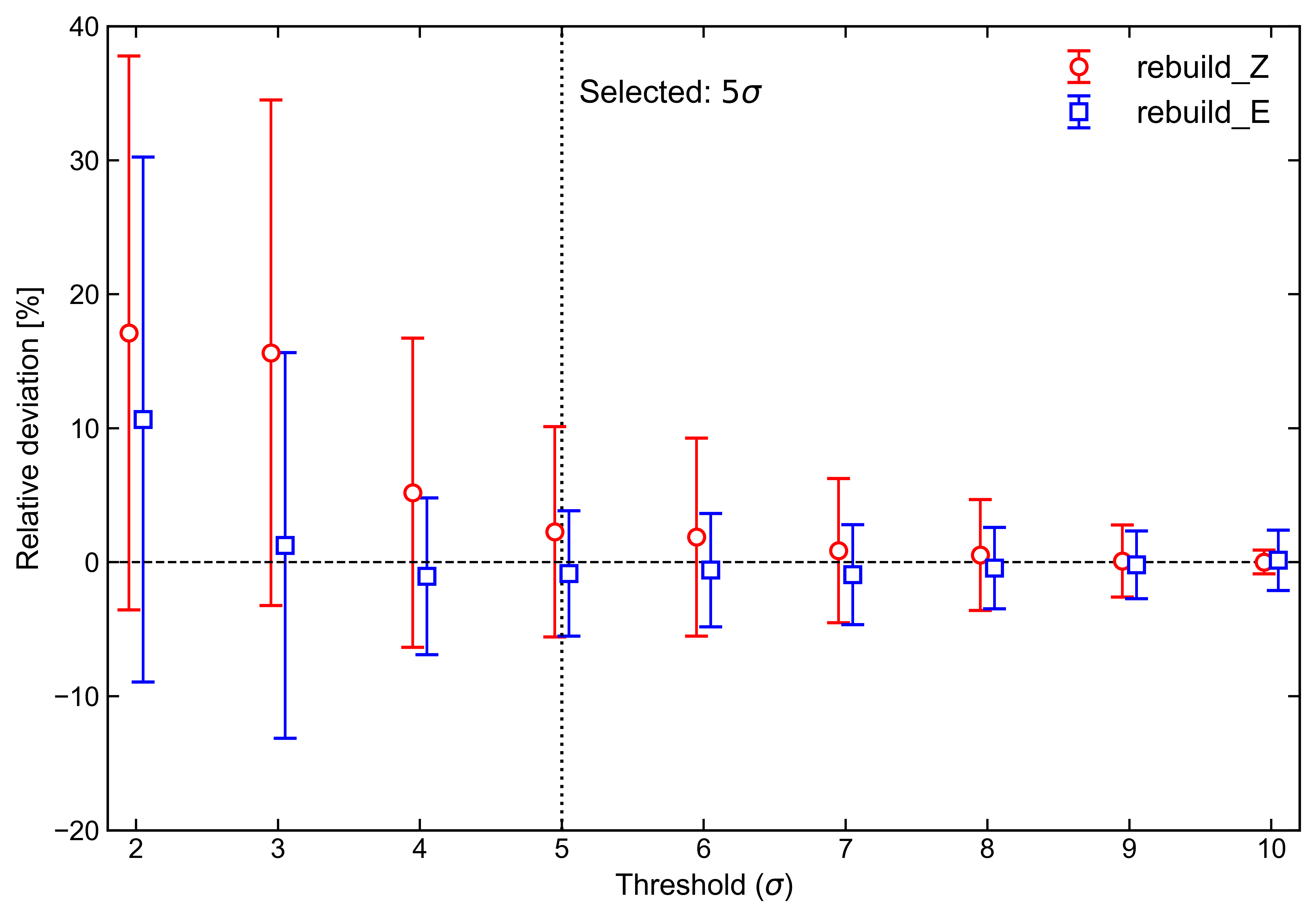}
    \caption{Reconstruction errors of the energy and drift distance for signals of different magnitudes.}
    \label{fig:rebuild_ZE_errorbar}
\end{figure}

To choose $S_t$, we assign different charge signals to 1000 channels, increasing the signal amplitude in units of the noise level. For each amplitude, we compare the reconstructed energy and drift distance with the true values. Figure~\ref{fig:rebuild_ZE_errorbar} shows the corresponding reconstruction deviations.

As expected, larger signals yield smaller deviations. When the signal exceeds five times the noise level, both the energy and drift-distance deviations are less than 10\%. We therefore set the signal threshold to $S_t = 5\sigma_{\mathrm{noise}} = 225~e^{-}$. Using the charge-to-voltage gain of the Topmetal-S sensor, this corresponds to approximately 7.5~mV. Signals below this threshold are discarded.

To further reduce the impact of noise, we apply a low-pass filter to the waveform before extracting the initial values for the fit. The choice of the cutoff frequency $f_c$ must balance two competing effects: a lower $f_c$ produces a smoother waveform, which improves the stability of the initial-value estimation, but an excessively low $f_c$ distorts the waveform and prevents accurate detection of the two signal peaks.

\begin{figure}[!htb]
    \centering
    \includegraphics[width=0.55\textwidth]{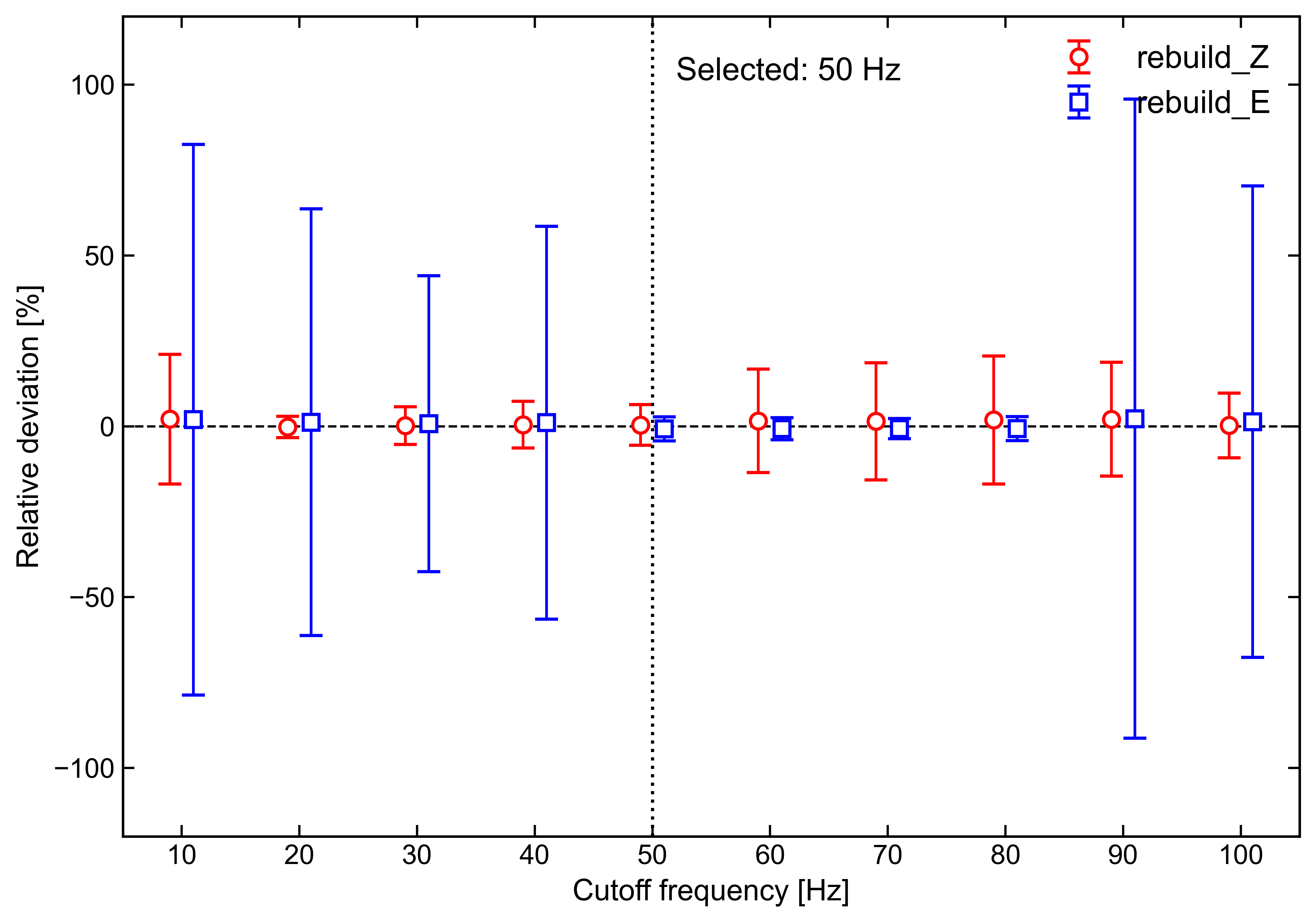}
    \caption{Reconstruction errors of the energy and drift distance for different low-pass filter cutoff frequencies.}
    \label{fig:rebuild_ZE_cutoff_frequency_errorbar}
\end{figure}

To determine a universal $f_c$ for signals with different charge amounts and drift distances, we simulate signals at a fixed noise level of 45~$e^{-}$ and apply low-pass filters with various cutoff frequencies. We then extract the initial parameters, perform the waveform fitting, and compare the reconstructed drift distances and channel charges with the true values. Figure~\ref{fig:rebuild_ZE_cutoff_frequency_errorbar} shows the reconstruction errors as functions of the cutoff frequency.

Both excessively high and low $f_c$ lead to significant deviations between the reconstructed and true values. Balancing the reconstruction accuracy of both the energy and drift distance, we select 50~Hz as the cutoff frequency. The final optimized parameters are summarized in Table~\ref{tab:optimized_parameters}.

\begin{table}[!htb]
\centering
\caption{Optimized parameter selection.}
\label{tab:optimized_parameters}
\begin{tabular}{lccc}
\toprule
Parameter & Value & Drift-distance error & Energy error \\
\midrule
$S_t$ & 7.5~mV & $<10\%$ & $<4\%$ \\
$f_c$ & 50~Hz & $<5\%$ & $<1\%$ \\
\bottomrule
\end{tabular}
\end{table}

\subsection{Reconstruction Results}

\begin{figure*}[!htb]
    \centering

    \begin{subfigure}[t]{0.30\textwidth}
        \centering
        \includegraphics[width=\textwidth]{diffusion_xy.png}
        \caption{}
        \label{fig:diffusion_track_xy2}
    \end{subfigure}
    \hfill
    \begin{subfigure}[t]{0.30\textwidth}
        \centering
        \includegraphics[width=\textwidth]{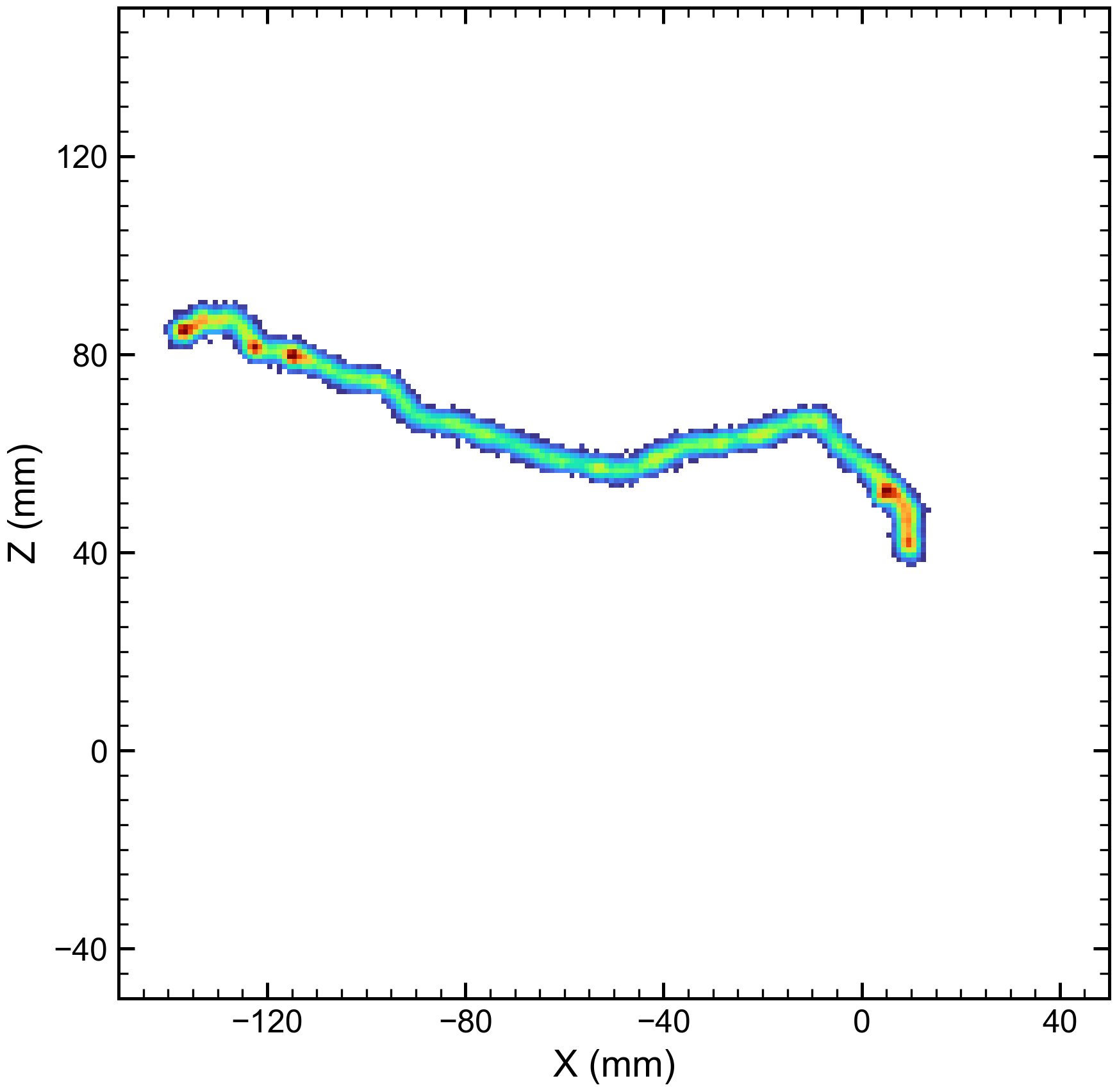}
        \caption{}
        \label{fig:diffusion_track_xz2}
    \end{subfigure}
    \hfill
    \begin{subfigure}[t]{0.30\textwidth}
        \centering
        \includegraphics[width=\textwidth]{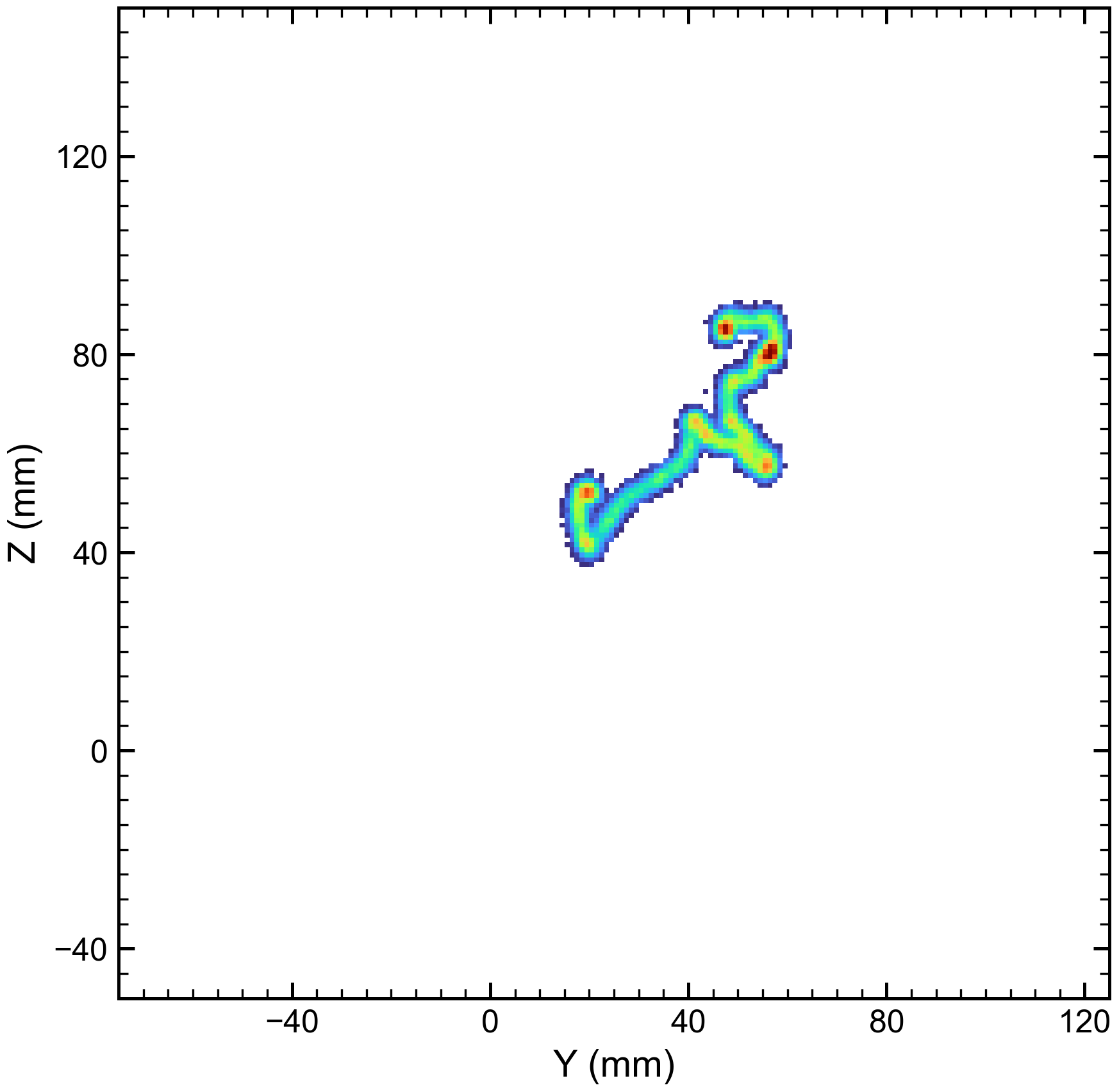}
        \caption{}
        \label{fig:diffusion_track_yz2}
    \end{subfigure}

    \vspace{0.3cm}

    \begin{subfigure}[t]{0.30\textwidth}
        \centering
        \includegraphics[width=\textwidth]{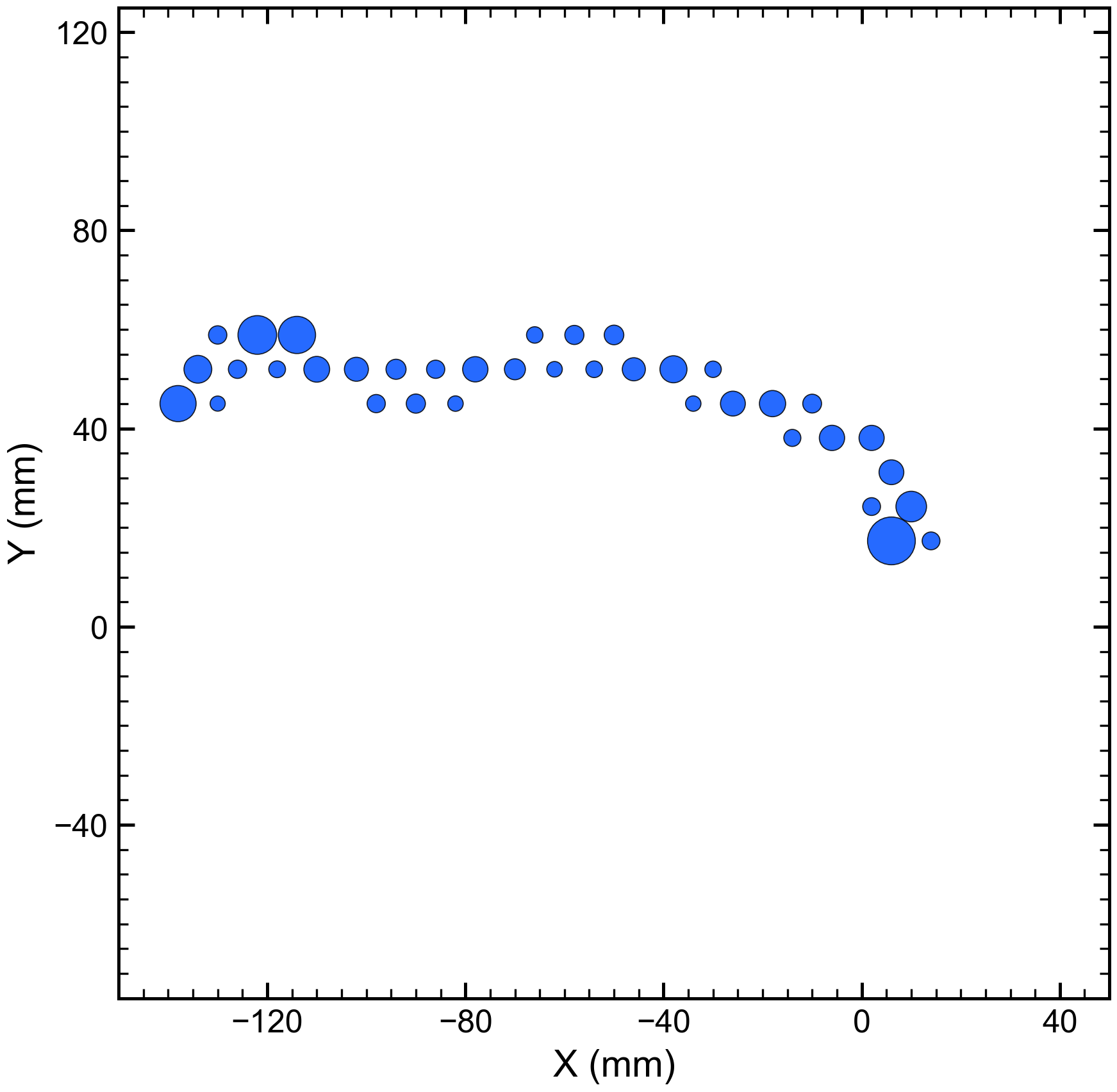}
        \caption{}
        \label{fig:rebuild_track_xy}
    \end{subfigure}
    \hfill
    \begin{subfigure}[t]{0.30\textwidth}
        \centering
        \includegraphics[width=\textwidth]{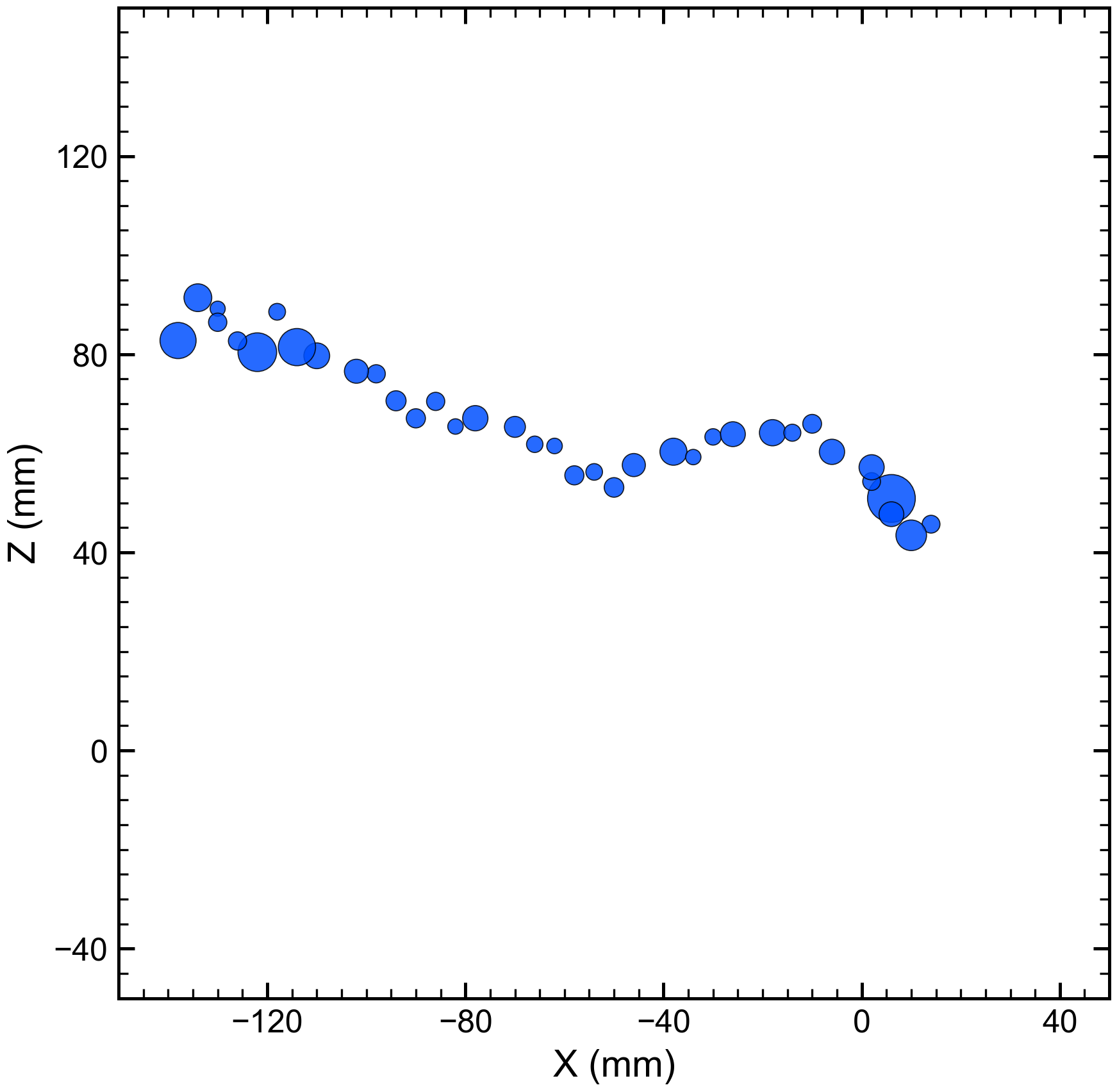}
        \caption{}
        \label{fig:rebuild_track_xz}
    \end{subfigure}
    \hfill
    \begin{subfigure}[t]{0.30\textwidth}
        \centering
        \includegraphics[width=\textwidth]{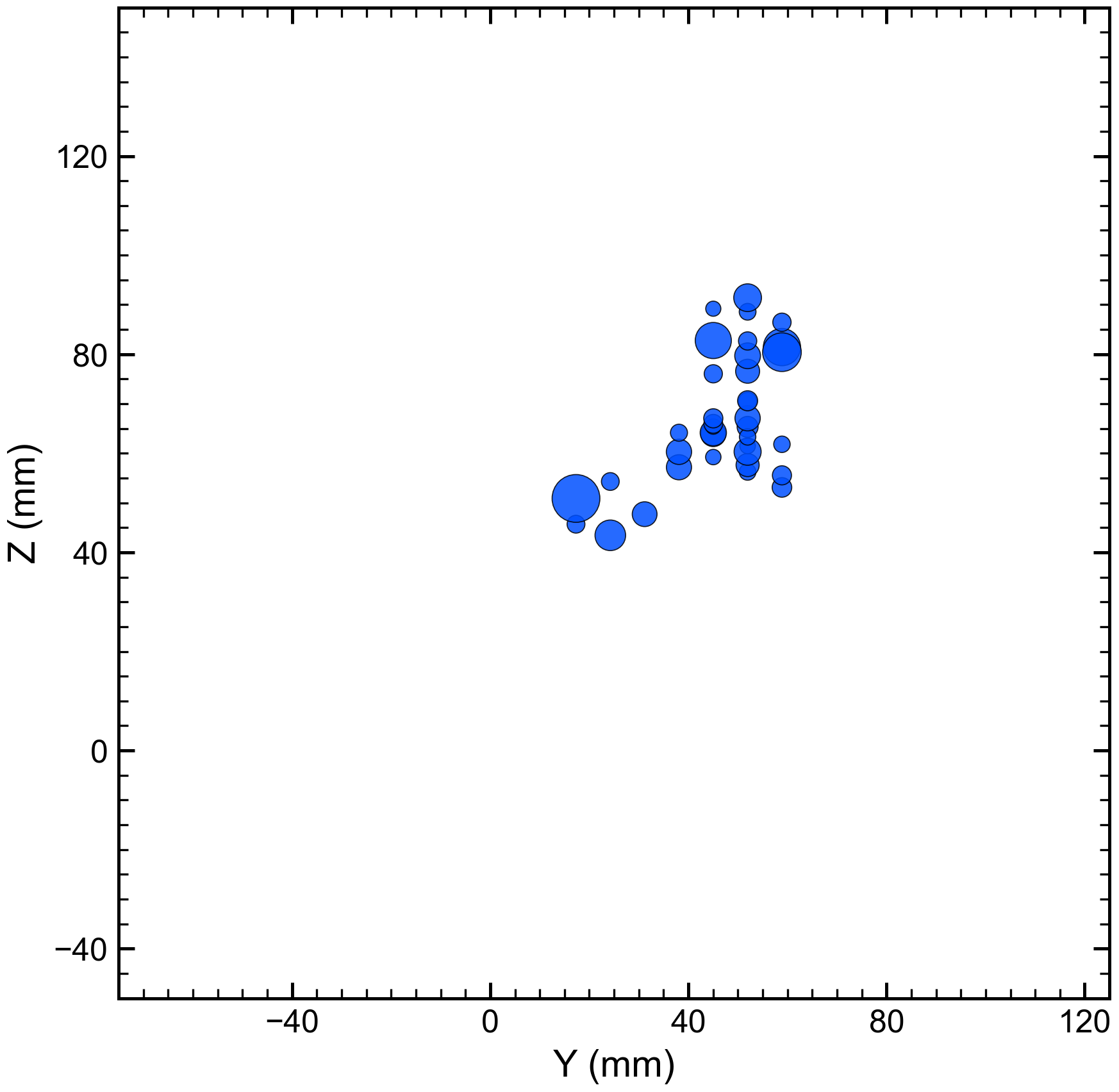}
        \caption{}
        \label{fig:rebuild_track_yz}
    \end{subfigure}

    \caption{Comparison between the diffused and reconstructed tracks in different projections.
    (a)--(c) Diffused tracks in the XY, XZ, and YZ projections, respectively.
    (d)--(f) Corresponding reconstructed tracks in the XY, XZ, and YZ projections, respectively.}
    \label{fig:rebuild_track}
\end{figure*}

After applying the reconstruction method with the optimized parameters, the distribution of ionization energy-loss points is reconstructed from the chip signals. Figure~\ref{fig:rebuild_track} compares the diffused and reconstructed tracks in the XY, XZ, and YZ projections.

Summing the reconstructed energy of each ionization point yields the total reconstructed event energy. For the $0\nu\beta\beta$ decay of $^{82}\mathrm{Se}$, we aim to minimize the energy resolution of the reconstructed spectrum. We fit the reconstructed energy spectrum with the Crystal Ball function:
\begin{equation}
f(x)=N\cdot
\begin{cases}
e^{-\frac{(x-\mu)^2}{2\sigma^2}}, & \displaystyle\frac{x-\mu}{\sigma}>-\alpha,\\[6pt]
A\cdot\left(B-\displaystyle\frac{x-\mu}{\sigma}\right)^{-n}, & \text{otherwise},
\end{cases}
\label{eq:CrystalBall}
\end{equation}
where
\begin{equation}
A=\left(\frac{n}{|\alpha|}\right)^n \exp\!\left(-\frac{|\alpha|^2}{2}\right),\quad
B=\frac{n}{|\alpha|}-|\alpha|.
\label{eq:CrystalBall_constants}
\end{equation}

\begin{figure}[!htb]
    \centering
    \includegraphics[width=0.55\textwidth]{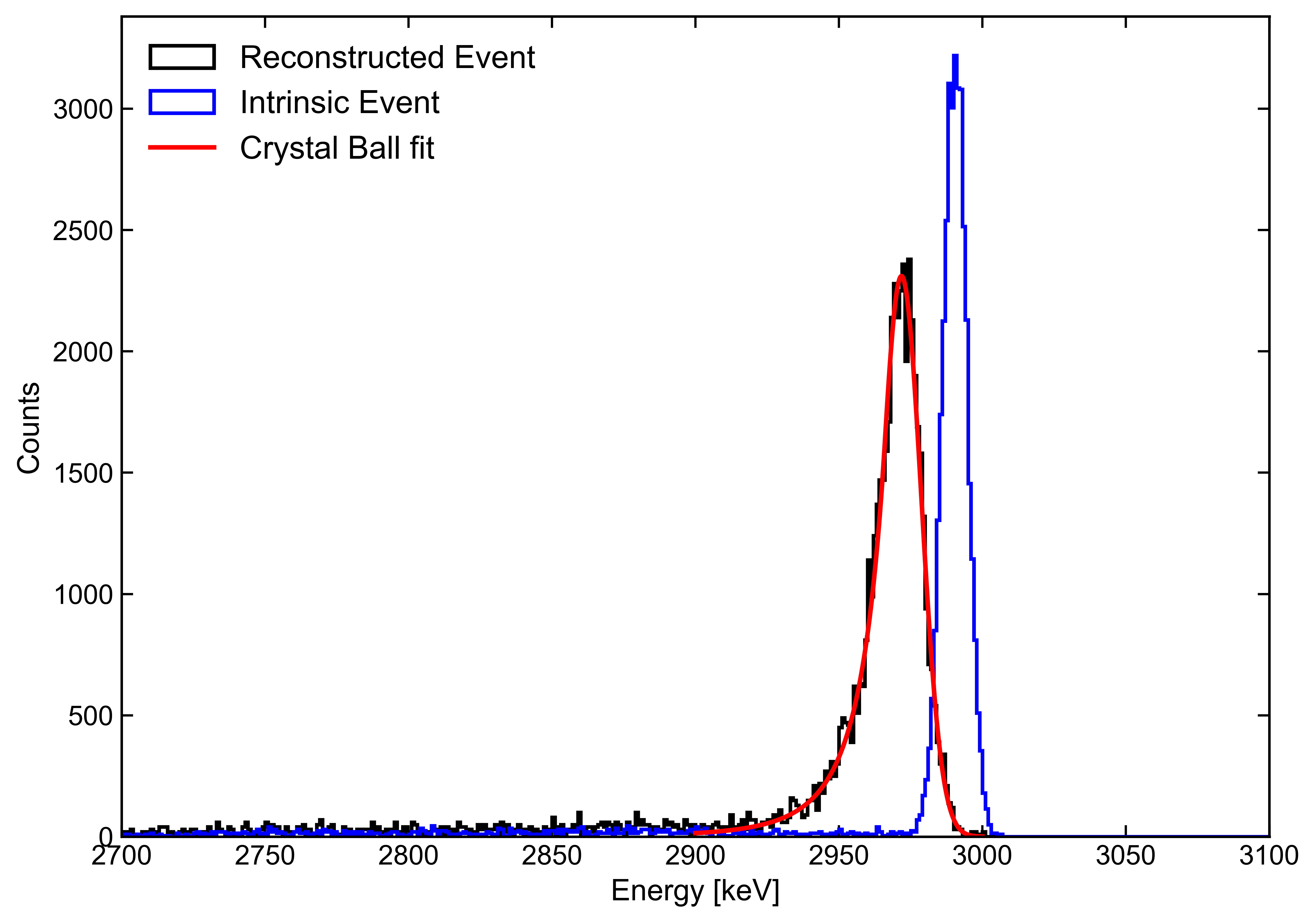}
    \caption{
    Comparison of the intrinsic event energy spectrum and the reconstructed energy spectrum. The intrinsic spectrum is fitted with a Gaussian function, while the reconstructed spectrum is fitted with a Crystal Ball function.
    }
    \label{fig:energy_spectrum_comparison}
\end{figure}

Figure~\ref{fig:energy_spectrum_comparison} shows the fitting result, with $\chi^2/\mathrm{ndf}=116/100$. The reconstructed FWHM energy resolution is
\begin{equation}
\mathrm{FWHM} =
\frac{7.288}{2971}\cdot 2\sqrt{2\ln 2}
\approx 0.58\%.
\label{eq:FWHM}
\end{equation}

Before reconstruction, a Gaussian fit to the intrinsic event energy spectrum yields $\chi^2/\mathrm{ndf}=146/100$ and the FWHM energy resolution of $0.33\%$. After reconstruction and parameter optimization, the reconstructed energy spectrum yields $\chi^2/\mathrm{ndf}=116/100$ and the FWHM energy resolution of $0.58\%$, which is consistent with the theoretical estimate of $0.61\%$ obtained in Appendix~\ref{app:energy_resolution}.

\section{Summary and Outlook}
\label{sec:conclusion}

In this work, we developed a full simulation and reconstruction framework for the N$\nu$DEx-100 experiment based on the REST toolkit. The simulation chain implements the detector geometry, the high-pressure $\mathrm{SeF_6}$ gas active volume, the Topmetal pixelated readout plane, negative-ion drift and diffusion, signal formation, electronics response, and noise contributions. The geometry model is constructed via GDML and imported into REST, providing a realistic detector description for particle transport and detector-response simulations.

A dedicated ion-diffusion process describes the charge transport in electronegative $\mathrm{SeF_6}$ gas. Because ionized electrons are quickly captured by gas molecules, the signal transport in the TPC is dominated by negative ions, primarily $\mathrm{SeF_5^-}$ and $\mathrm{SeF_6^-}$. The different drift velocities of these species produce distinguishable double-peak waveforms on the readout channels, providing timing information that allows the drift distance of the ionization point to be reconstructed. The Topmetal pixelated readout response is also modeled, converting the diffused charge distribution into activated pixels and channel waveforms.

Based on the simulated chip signals, we developed a waveform-fitting reconstruction method. The fitting function combines the double-peak ion-arrival structure with the chip's electronics response, enabling the extraction of peak times and amplitudes from each signal waveform. The time difference between the two peaks is used to determine the drift distance, while the fitted amplitudes give the collected charge and deposited energy. With the optimized reconstruction parameters, the track topology and event energy are recovered from the simulated readout signals.

The reconstructed energy spectrum yields a FWHM energy resolution of $0.58\%$ at the $Q_{\beta\beta}$ region, satisfying the N$\nu$DEx-100 design requirement of $<1\%$. This result demonstrates that our simulation and reconstruction chain successfully reproduces the main detector-response features relevant to energy reconstruction. The framework also provides a foundation for further studies, including background rejection, event-topology analysis, readout optimization, and reconstruction-algorithm development.

Several directions for future refinement are apparent. The simulation chain can incorporate more detailed detector conditions, such as nonuniform electric fields, improved electronics response models, and realistic background sources. The reconstructed event information produced by this framework will also support topological event classification and sensitivity projections for the N$\nu$DEx-100 experiment.

\bibliographystyle{unsrt}
\bibliography{areference}

\vspace{1cm}

\appendix

\section{Energy Resolution: Theoretical Calculation}
\label{app:energy_resolution}

The $Q_{\beta\beta}$ value of $^{82}\mathrm{Se}$ is 2996~keV, the average ionization energy of $\mathrm{SeF_6}$ is $W = 32$~eV, and the Fano factor is $F = 0.19$. The ratio of the standard deviation $\sigma_N$ to the total number $N$ of produced ions is
\begin{equation}
\frac{\sigma_N}{N}
= \frac{\sqrt{F\cdot Q_{\beta\beta}/W}}{Q_{\beta\beta}/W}
\approx 0.14\%.
\label{eq:sigma_n_N_app}
\end{equation}

For a Gaussian distribution, the FWHM is related to the standard deviation $\sigma$ by $\mathrm{FWHM} = 2\sqrt{2\ln 2}\,\sigma$. The Fano-limited FWHM energy resolution is therefore
\begin{equation}
\mathrm{FWHM}_1 = 0.14\%\cdot 2\sqrt{2\ln 2} \approx 0.33\%.
\label{eq:FWHM1_app}
\end{equation}

Assuming a chip noise of $a = 45$ electrons and an average of $b = 20$ chips carrying signals per event, the contribution from chip noise to the energy resolution is estimated as
\begin{equation}
\mathrm{FWHM}_2 = \frac{\sqrt{b}\cdot a}{Q_{\beta\beta}/W}\cdot 2\sqrt{2\ln 2} \approx 0.51\%.
\label{eq:FWHM2_app}
\end{equation}

The total energy resolution, combining the Fano and noise contributions in quadrature, is
\begin{equation}
\mathrm{FWHM}_{\mathrm{total}} = \sqrt{0.33^2 + 0.51^2}\,\% = 0.61\%.
\label{eq:FWHMtotal_app}
\end{equation}

The value $b = 20$ is representative of the typical number of readout pixels that register signals above threshold for the simulated $2\nu\beta\beta$ events in the N$\nu$DEx-100 detector. This estimate provides a first-order approximation of the noise contribution; the Monte Carlo simulation implicitly accounts for the actual pixel multiplicity distribution.

\section{Derivation of the Ion Mobility Mass-Scaling Relation}
\label{app:mobility_scaling}

For ion species without direct mobility measurements, we estimate the reduced mobility using an approximate mass-scaling relation derived from a simplified collision model. In this model, the mass dependence of the ion drift velocity follows from the ion acceleration in the electric field and the mean free time between ion--molecule collisions.

For a particle in thermal equilibrium, the average thermal kinetic energy is approximately independent of particle mass at a fixed temperature. Thus, the thermal velocity scales as $v_T \propto 1/\sqrt{A}$, where $A$ is the mass number of the particle. For collisions between a drifting ion and a gas molecule, the relative thermal velocity is
\begin{equation}
v_{\mathrm{rel}} \propto \sqrt{v_{T,\mathrm{ion}}^2 + v_{T,\mathrm{molecule}}^2}
\propto \sqrt{\frac{1}{A_{\mathrm{ion}}} + \frac{1}{A_{\mathrm{molecule}}}}.
\end{equation}

If the collision cross sections and gas density are approximately equal for the ion species considered, the mean free time between collisions is inversely proportional to the relative velocity:
\begin{equation}
\tau \propto \frac{1}{v_{\mathrm{rel}}}
\propto \frac{1}{\sqrt{1/A_{\mathrm{ion}} + 1/A_{\mathrm{molecule}}}}.
\end{equation}

Under an electric field, the acceleration of a singly charged ion is $a = qE/m_{\mathrm{ion}} \propto 1/A_{\mathrm{ion}}$. In a simplified free-flight picture, the average drift velocity over one mean free time $\tau$ is
\begin{equation}
v_d \simeq \frac{1}{2} a\tau \propto a\tau.
\end{equation}

Combining the mass dependence of $a$ and $\tau$,
\begin{equation}
v_d \propto \frac{1/A_{\mathrm{ion}}}{\sqrt{1/A_{\mathrm{ion}} + 1/A_{\mathrm{molecule}}}}.
\end{equation}

Because the mobility is $\mu = v_d/E$, the same mass dependence applies to the reduced mobility $\mu_0$ under comparable gas conditions. Taking the measured reduced mobility of $\mathrm{SF_6^-}$ in $\mathrm{SF_6}$ gas as a reference, the reduced mobility of an ion with mass number $A_{\mathrm{ion}}$ in $\mathrm{SeF_6}$ gas is estimated as
\begin{equation}
\mu_0(A_{\mathrm{ion}}) =
\mu_0(\mathrm{SF_6^-})
\frac{
\displaystyle \frac{1/A_{\mathrm{ion}}}{\sqrt{1/A_{\mathrm{ion}}+1/A_{\mathrm{SeF_6}}}}
}{
\displaystyle \frac{1/A_{\mathrm{SF_6}}}{\sqrt{1/A_{\mathrm{SF_6}}+1/A_{\mathrm{SF_6}}}}
}.
\label{eq:mu_scaling}
\end{equation}

Here, $A_{\mathrm{SF_6}} = 146$ and $A_{\mathrm{SeF_6}} = 196$ are the mass numbers of the $\mathrm{SF_6}$ and $\mathrm{SeF_6}$ molecules, respectively.

\end{document}